\documentclass{aa} % for a referee version
\usepackage{ulem}
\usepackage{graphicx}
\usepackage{url}
\usepackage{color}
\usepackage{xcolor}
\definecolor{xlinkcolor}{cmyk}{1,1,0,0}
\usepackage{threeparttable}
\usepackage[whole]{bxcjkjatype}
\usepackage{enumerate}
\usepackage{physics}
\usepackage{threeparttable}
\usepackage{txfonts}
\newcommand{\rmf}[1]{{_{\rm #1}}}
\begin{document}

   \title{Simulations of two-temperature jets in galaxy clusters}

   \subtitle{ II. X-ray property of forward shock}

   \author{T. Ohmura\inst{1,2,3}
          \and
          M. Machida\inst{3}
          \and
          H. Akamatsu\inst{4,5}
          }

   \institute{Institute for Cosmic Ray Research, The University of Tokyo, 5-1-5 Kashiwanoha, Kashiwa, Chiba 277-8582 Japan\\
              \email{tohmura@icrr.u-tokyo.ac.jp}
         \and
         Department of Physics, Faculty of Sciences, Kyushu University, 744 Motooka, Nishi-ku, Fukuoka 819-0395, Japan\\
         \and
        Division of Science, National Astronomical Observatory of Japan, 2-21-1 Osawa, Mitaka, Tokyo 181-8588, Japan\\
             \email{mami.machida@nao.ac.jp}
         \and
        SRON Netherlands Institute for Space Research, Niels Bohrweg 4, 2333 CA Leiden, the Netherlands\\
             \email{h.akamatsu@sron.nl}  
         \and
         International Center for Quantum-field Measurement Systems for Studies of the Universe and Particles,  the High Energy Accelerator Research Organization, Tsukuba, Ibaraki, Japan\\}

   \date{accepted September 11, 2023}

% \abstract{}{}{}{}{} 
% 5 {} token are mandatory
 
  \abstract
   { Forward shocks by radio jets, driven into the intracluster medium, are one of the indicators that can be used to evaluate the power of the jet. Meanwhile high-angular-resolution X-ray observations show the Mach numbers of powerful radio jets are smaller compared to that of theoretical and numerical studies, $\mathcal{M_{\rm obs}} < 2$.  }
%   {We systematically investigate various effects that influence the estimates of the Mach number in powerful jets.}
    {
    Our aim is to systematically investigate various factors, such as projection effects and temperature non-equilibration between protons and electrons, that influence the Mach number estimate in a powerful jet.
    }
   {Using two-temperature magnetohydrodynamic simulation data for the Cygnus A radio jets, whose Mach number is approximately 6, we construct mock X-ray maps of simulated jets from various viewing angles. Further, we evaluate the shock Mach number from density/temperature jump using the same method of X-ray observations. }
   {Our results demonstrate that measurements from density jump significantly underestimate the Mach numbers, $\mathcal{M} < 2$, around the jet head at a low viewing angle, $\lessapprox 50^{\circ}$.
   The observed post-shock temperature is strongly reduced by the projection effect, as our jet is in the cluster center where the gas density is high.
   On the other hand, the temperature jump is almost unity, even if thermal electrons are in instant equilibration with protons. Upon comparison, we find that shock property of our model at viewing angle of $<$ 55º is in a good agreement with that of Cygnus A observations.
   }
  {
    These works illustrate that the importance of the projection effect to estimate the Mach number from the surface brightness profile.
    Furthermore, forward shock Mach numbers could be a useful probe to determine viewing angles for young, powerful radio jets.
  }


   \maketitle
%
%-------------------------------------------------------------------
\section{Introduction}
\label{sec:intro}
Powerful radio jets are launched from radio-loud active galactic nuclei (AGN) \citep[see review by][]{2019ARA&A..57..467B}.
AGN jets can be grouped into two categories at kilo--parsec scales: the Fanaroff \& Riey (FR) class I and FR class II sources \citep{1974MNRAS.167P..31F}.
Typical FR II sources are more powerful than FR I sources, and the jet beam of the FR II source seems to maintain a relativistic velocity until its termination.
The kinetic energy of these jets is converted to heating energy in the intracluster medium (ICM) through shocks, and it is widely accepted that the radio-mode feedback plays a fundamental role in the formation and evolution of galaxies and large-scale structure \citep[e.g.,][and references therein]{2006PhR...427....1P,2007ARA&A..45..117M,2012ARA&A..50..455F}.
However, there are many remaining questions about the energetics and dynamical properties of AGN jets. 

Theoretical models of the FR II type radio source are well established \citep{1974MNRAS.166..513S,1974MNRAS.169..395B,1997MNRAS.286..215K}.
\citet{1989ApJ...345L..21B} showed that accumulated plasma in the lobe is highly over-pressured against the ICM, and hence provides a strong forward shock drive into it.
Therefore, a high density shell can be formed around the lobe.
This is in agreement with hydrodynamic and magnetohydrodynamic (MHD) simulations of jet propagation 
\citep[e.g.,][]{2003A&A...398..113K,2009MNRAS.400.1785G,2013MNRAS.430..174H,2014MNRAS.445.1462P}
In particular, when the jet beam can propagate stably and reach the jet head, the Mach number around the jet head is high \citep[e.g.,][]{2018MNRAS.481.2878E,2019MNRAS.482.3718P}.

In high-angular-resolution X-ray observations, forward shocks are observed for both FR I and FR II sources.
For example, the discontinuity in the X-ray image due to the forward shocks is clearly visible in Hydra A \citep{2005ApJ...628..629N}, MS0735.6+7421 \citep{2005Natur.433...45M}, and Cygnus A \citep{2006ApJ...644L...9W}.
Furthermore, ripples, like weak shocks, are observed around the radio lobes for the Perseus cluster \citep{2006MNRAS.366..417F} and M87 \citep{2007ApJ...665.1057F}. 
The forward shock information provide important clues to understand the energetics of radio jets and the nature of the tenuous plasma.
furthermore, the forward shocks would be expected to be a possible cosmic-ray acceleration site \citep{2007ApJ...663L..61F,2011ApJ...730..120I}.
In fact, \citet{2009MNRAS.395.1999C} reported the detection of non-thermal X-ray components from the forward shock associated with Centaurus A. 

Using deep {\it Chandra} X-ray data, \citet{2018ApJ...855...71S} investigated the properties of forward shocks for Cygnus A, that is the archetype of Fanaroff-Riley type II radio galaxies \citep{1974MNRAS.167P..31F,1996A&ARv...7....1C}. 
The estimated jet power of Cygnus A is in the range of $10^{45}-10^{46}$ $\rm{erg \, s^{-1}}$.
Surprisingly, the measured Mach numbers of the forward shocks of Cygnus A are below two, even around the jet head.
This is inconsistent with several analytical and numerical studies.
Meanwhile, \citet{2017MNRAS.467.1586I} performed another estimation of the shock Mach number for a large sample of the FR II radio galaxies lobes, at redshifts of 0.1 and 0.5.
Using non-thermal X-ray emission data, they evaluated the internal pressure of relativistic electrons in the lobe, and derived the Mach number from the pressure ratio between internal lobe and external ICM.
Their results suggest that the median value of the Mach number is about two.
However, this value might be somewhat underestimated, as it does not include the contribution of non-radiation particles in a lobe. 

The forward shock of AGN jet would be an ideal celestial laboratory to examine the electron heating mechanism of collisionless shock.
In a collisionless system, the efficient heating of electrons is not trivial \citep{doi:10.1029/JA093iA11p12923,2010PhPl...17d2901M,2015A&A...579A..13V,2017ApJ...851..134G,2018ApJ...858...95G,2020ApJ...900L..36T}.
Heavier protons hold most of the bulk kinetic energy, and hence retain most of the thermal energy at the downstream shock.
Furthermore, the timescale of the proton-electron temperature equilibrium is given by
\begin{equation}
\label{eq:teq}
t_{\rm ep} = 20~{\rm Myr} \left( \frac{\ln{\Lambda}}{40}\right)^{-1} \left( \frac{n_{\rm p}
}{10^{-2} {\rm cm^{-3}}}\right)^{-1} \left( \frac{T_{\rm e}}{10^8 {\rm K}} \right)^{3/2} , 
\end{equation}
 which is comparable to the shock propagation time scale, as the ICM is hot and has low density \citep{1998ApJ...509..579T}.
This is, therefore, expected to be the timescale of the temperature equilibrium in the far downstream of the forward shock.

In the context of the X-ray observations of galaxy clusters, the shock Mach number can be measured independently using the X-ray surface brightness and spectroscopic temperature \citep{2007PhR...443....1M}.
Some observations indicate the existence of a temperature non-equilibrium in the post-shock region  \citep{2010arXiv1010.3660M,2010PASJ...62..371H,2011PASJ...63S1019A,2018ApJ...856..162W}
Both the projection and viewing angle affect the measured Mach number.
Several studies of cluster shocks examined and discussed these effects \citep{2013ApJ...765...21S,2015ApJ...812...49H,2019MNRAS.482...20Z,2020MNRAS.495.5014B,2021MNRAS.506..396W}, however, for the forward shock of AGN jets, no quantitative discussion of this effect has been presented so far.
Furthermore, when measuring the shock Mach number from X-ray surface brightness profile, the actual density profiles are obtained by model fitting \citep[e.g.,][]{2009ApJ...704.1349O}.
Hence, we must address this model dependence.

We studied the dynamical properties of powerful jets propagating within the Cygnus-A like cluster using two-temperature MHD simulations presented in Ohmura et al. (2022, hereafter referred to as Paper I).
The aim of this study is to systematically investigate various effects, such as projection and temperature non-equilibrium, that influence the estimates of the Mach number from X-ray surface brightness profiles and spectroscopic temperature jumps in powerful radio jets.
To achieve this purpose, we first focus on the thermodynamics of the shocked-ICM, which is heated by the forward shock.
Then, we conduct mock X-ray observations at several viewing angles to investigate the thermal X-ray properties of the forward shock.

Our paper is structured as follows: in Section \ref{sec:method}, we introduce the model for our MHD simulations and the numerical method of mock X-ray observations.
We report our MHD simulation results for the thermodynamical properties of the shocked-ICM and evolution of the forward shock.
Section \ref{sec:mockX} describes the results of mock X-ray observations in terms of their dependence on the viewing angle and the fitting model for X-ray surface brightness (section \ref{sec:mockX-sb}) and spectroscopic-like temperature (Section \ref{sec:mockX-tspec}).
The model dependence of the fitting results is discussed using actual X-ray data of Cygnus A in \ref{sec:cygAshock}. 
Finally, summary and discussions are given in section \ref{sec:summary}.

%--------------------------------------------------------------------
\section{Mock X-ray observation}
\label{sec:method}

\subsection{Numerical model}
\label{sec:method-model}
We presented the results of the two-temperature MHD jets in Paper I using CANS+ MHD code \citep{2019PASJ...71...83M}.
In this study, we analyzed only the model B jet in Paper I.
We briefly summarize the model and method employed in our MHD simulations.
The simulations were carried out in a Cartesian domain of size $(L_x, L_y, L_z) \in (\pm 32, \pm 32, 96)$ kpc.
The kinetic power of our jet and ICM properties are roughly consistent with Cygnus A \citep{2013ApJ...767...12G}.
Important parameters for the jet are listed in Table \ref{tab:model}.
The density profile of the ICM is given by
\begin{equation}
  \label{eq:beta}
  n(r) = \frac{n_0}{\left[ 1+(r/r\rmf{c})^2 \right]^{3\beta/2}},
\end{equation}
where $r=\sqrt{x^2+y^2+z^2}$, $n_0$, $r\rmf{c}$, and $\beta$ represent the radius, core density, core radius, and ratio of the specific energy in the galaxies to the specific thermal energy in the ICM, respectively \citep{1962AJ.....67..471K,2002ApJ...565..195S,2006ApJ...644L...9W}.
We set $\beta = 0.5$, $r\rmf{c} = 20$ kpc, and $n_0 =$ 0.05 ${\rm cm^{-3}}$.
Furthermore, our atmosphere is initially isothermal with a temperature $kT\rmf{p} = kT\rmf{e} = 5$ keV.

Our simulations implemented a temperature non-equilibrium between the electron and proton.
To calculate two-temperature plasma, we determine both the proton- and electron-specific entropies.
Electrons and protons exchange energy by Coulomb collisions, and electrons lose energy by thermal free-free emission \citep{1962pfig.book.....S, 1983MNRAS.204.1269S}.
We assume that electrons can receive 5 \% of the dissipated energy as shock waves.
In areas other than in the shock regions, the dissipated energy is divided by the protons and electrons using the sub-grid model for the turbulent damping process, which is derived from gyrokinetic simulations \citep{2019PNAS..116..771K}.
Details on the numerical method and model are provided in Paper I.

%  -------------------------- Table -------------------------  %
\begin{table}
  \begin{center}
  \caption{Jets and ICM setup parameters of MHD simulation}
  \label{tab:model}
    \begin{tabular}{c c c}
      \hline\hline
      Jet speed               & $v_{\rm jet}$     & 0.3$c$ \\
      Jet gas temperature     & $T_{\rm g,jet}$   & $10^{10}$ K \\
      Jet Kinetic power       & $L_{\rm kin}$     & $5.0\times10^{45}$ erg ${\rm s^{-1}}$ \\
      Jet thermal power       & $L_{\rm th}$      & $4.4\times10^{44}$ erg ${\rm s^{-1}}$ \\
      Jet radius              & $r_{\rm jet}$     & 1 kpc \\
      Jet Sonic Mach Number   & ${\mathcal M}_{\rm jet}$ & 6.2 \\ 
      Jet magnetic field      & $B_{\rm \phi,jet}$ & 6.17 $\mu {\rm G}$ \\
      Jet plasma beta         & $\beta_{\rm jet}$ & 5  \\ \hline
      ICM temperature         & $T_{\rm ICM}$     & 5 keV \\
      Core density            & $  n_{0}$         & $5\times10^{-2}~{\rm cm^{-3}}$ \\
      Core radius             & $r\rmf{c}$        & 20 kpc \\
      Core parameter          & $\beta$           & 0.5 \\
      ICM magnetic field      & $B_{\rm z, ICM}$  & 0.44 $\mu G$     \\
      ICM plasma beta         & $\beta_{\rm gas,ICM}$ & 1000  \\ \hline
      Numerical domain        & \multicolumn{2}{c}{$(L_x, L_y, L_z) \in (\pm 32, \pm 32, 96)$ kpc}  \\
      Resolution              & \multicolumn{2}{c}{$640~\times~640~\times~960$  (Uniform grids)} \\
      \hline\hline
  \end{tabular}
\end{center}
\end{table}
%-------------------------------------------------------------------

\subsection{Mock X-ray observation}
\label{sec:method-xray}
X-ray observations provide the surface brightness and spectroscopic temperature obtained by fitting a thermal model to the X-ray spectrum.
X-ray surface brightness can be obtained by integral of the X-ray emissivity, which is summed of continuous and line components, from a optically thin plasma along the line of sight (LOS):
\begin{equation}
\label{eq:SB0}
{\rm SB} = \int_{\rm LOS} n^2_{\rm e} \Lambda(Z, T_{\rm e}) \dd V,
\end{equation}
where $\Lambda (Z,T_{\rm e})$, $Z$ and $dV$ are the cooling curve, metalicity, and the volume element, respectively.
The cooling curve $\Lambda (Z, T_{\rm e})$ for each element in specified spectral ranges is calculated as
\begin{equation}
    \label{eq:coolingcurve}
    \Lambda (Z,T_{\rm e}) =  \int \varepsilon_{\nu} (Z,T_{\rm e}) \dd \nu,
\end{equation}
where $\varepsilon_{\nu}$ is specific X-ray emissivity. 
To make cooling curves, we use a spectral model based on the APEC model for a thermal plasma from AtomDB database, and set a constant metalicity $Z = 0.5 Z_{\odot}$, which is suggested by the X-ray observation of Cygnus A \citep{2018ApJ...855...71S}

\begin{figure}
\centering
\includegraphics[width=0.95\columnwidth]{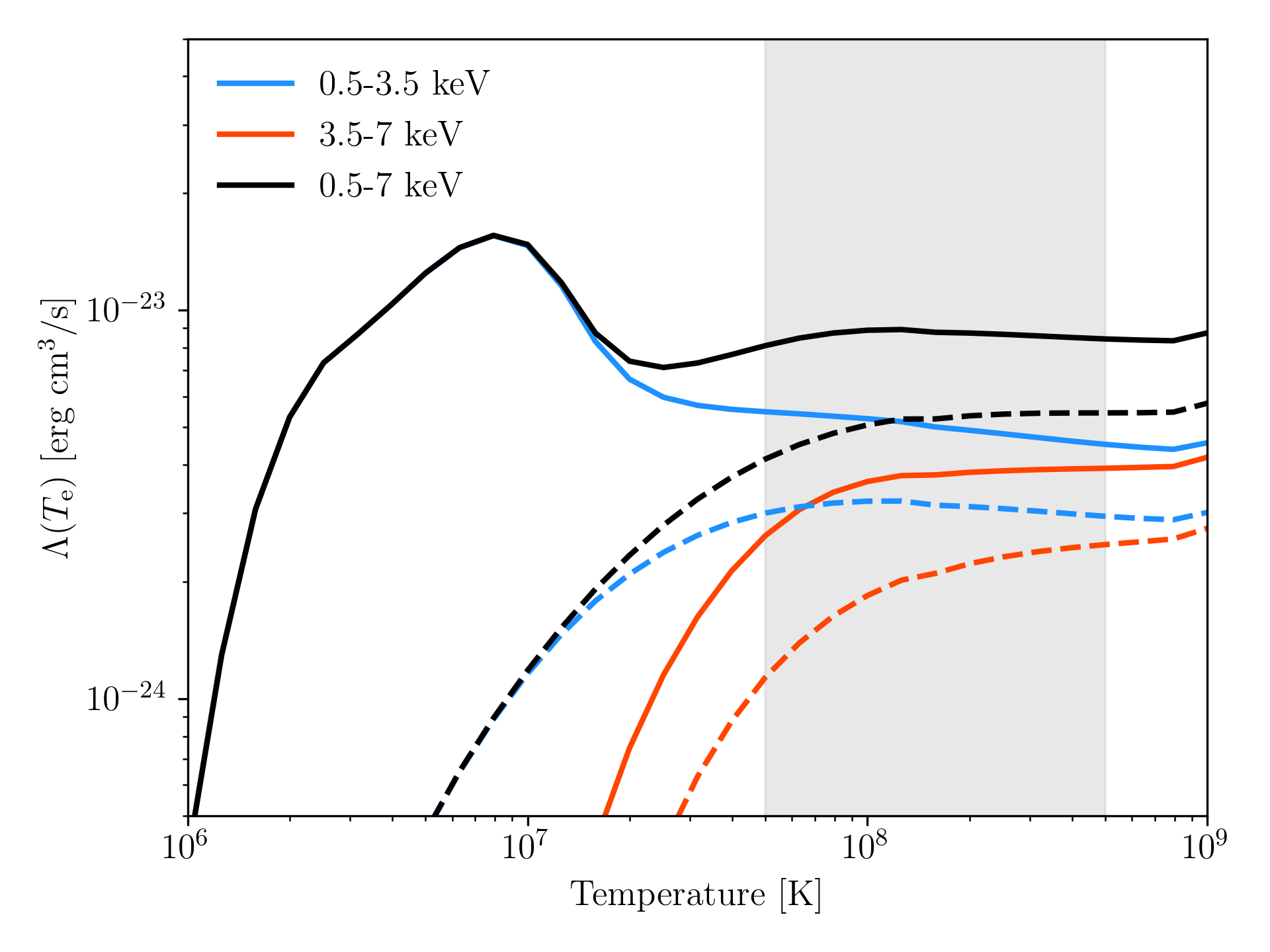}
\caption{Cooling curves in the energy bands: 0.5--3.5 keV (blue), 3.5--7 keV (red), and 0.5--7.0 keV (black). The dashed lines show only the bremssttrahulng emissions. We set $Z = 0.5 Z_{\odot}$. Shaded region shows a range of gas temperatures in the shocked-ICM of our simulations.   
}
    \label{fig1}%
\end{figure}

Figure \ref{fig1} shows cooling curves in the energy bands: 0.5--3.5 keV, 3.5--7 keV, and 0.5--7.0 keV. 
Cooling curves for 0.5 -- 7.0 keV (solid black) and 0.5 -- 3.5 keV (solid blue) have a very weak dependence on the electron temperature in a range of typical electron temperature of ICM.
Several studies of X-ray observations for galaxy clusters therefore assume that X-ray surface brightness can simply be obtained as follows \citep[e.g.,][]{2009ApJ...704.1349O, 2015MNRAS.450.4184Z, 2018ApJ...855...71S},
\begin{equation}
\label{eq:SB}
{\rm SB} = \int_{\rm LOS} n^2_{\rm e} \Lambda(T_{\rm e}) \dd V \approx A_0 \int_{\rm LOS} n_{\rm e}^2 dV,
\end{equation}
where $A_0$ is an constant.

For the spectroscopic temperature $T_{\rm spec}$, we adopt the spectroscopic-like formula proposed by \citet{2004MNRAS.354...10M}:
\begin{equation}
    \label{eq:tspec}
    T_{\rm spec} = \frac{\int_{\rm LOS} W T_{\rm e} dV }{ \int_{\rm LOS} W dV}, 
\end{equation}
where $T_{\rm e}$ is the electron temperature, and $W = n_{\rm e}^2 T^{-3/4}_{\rm e}$.
Non-thermal components contribute to the X-ray emission, especially in radio lobes.
In particular, non-thermal X-ray radiation from lobes is brighter than thermal radiation that comes from the ICM in the high-$z$ galaxy and/or powerful radio galaxy \citep{2020MNRAS.493.5181T}.
This study focuses on the nearby radio galaxy where the X-ray discontinuity is clearly visible.
Therefore, we neglect the non-thermal X-ray radiation in this study.
We use mock X-ray observation to calculate equations \eqref{eq:SB} and \eqref{eq:tspec} at $\pm 1$ Mpc along the LOS using snapshot data of our simulation.
This integral length is comparable with $r_{500}$ of Cygnus A \citep{2019MNRAS.483.3851H}. 
While our numerical model does not cover whole range of the cluster, we calculate the emissivity and spectroscopic-like temperature in the no-data region using extrapolated values from the $\beta$-profile (see equation \eqref{eq:beta}).

\subsection{Broken Power-law model of surface brightness profiles}
\label{sec:method-fit}
Density jumps at the discontinuity of X-ray surface brightness are conventionally determined by fitting with a broken power-law model, hereafter referred to as {\tt BknPow} \citep[e.g.,][]{2009ApJ...704.1349O}.
In this model, the electron density of the broken power-law model that assumes spherical symmetry is given by
\begin{equation}
    n_{\rm e}(r) = \begin{cases}
        n_0 \left( \frac{r}{r_{\rm sh}}\right)^{- \alpha_1} & (r < r_{\rm sh}) \\
        \frac{n_0}{C} \left( \frac{r}{r_{\rm sh}}\right)^{- \alpha_2} & (r > r_{\rm sh}),   
    \end{cases}
\end{equation}
where $n_0$ is the density normalization, $\alpha_1$ and $\alpha_2$ are the power-law indices, and $r_{\rm sh}$ is the shock radius.
At the density discontinuity location, $C$ is the shock compression factor, $C \equiv n_2/n_1$, where $n_1$ and $n_2$ are the pre- and post-shock densities, respectively.

%%------------------------------------------------------------%%
\begin{figure}
\centering
\includegraphics[width=0.9\columnwidth]{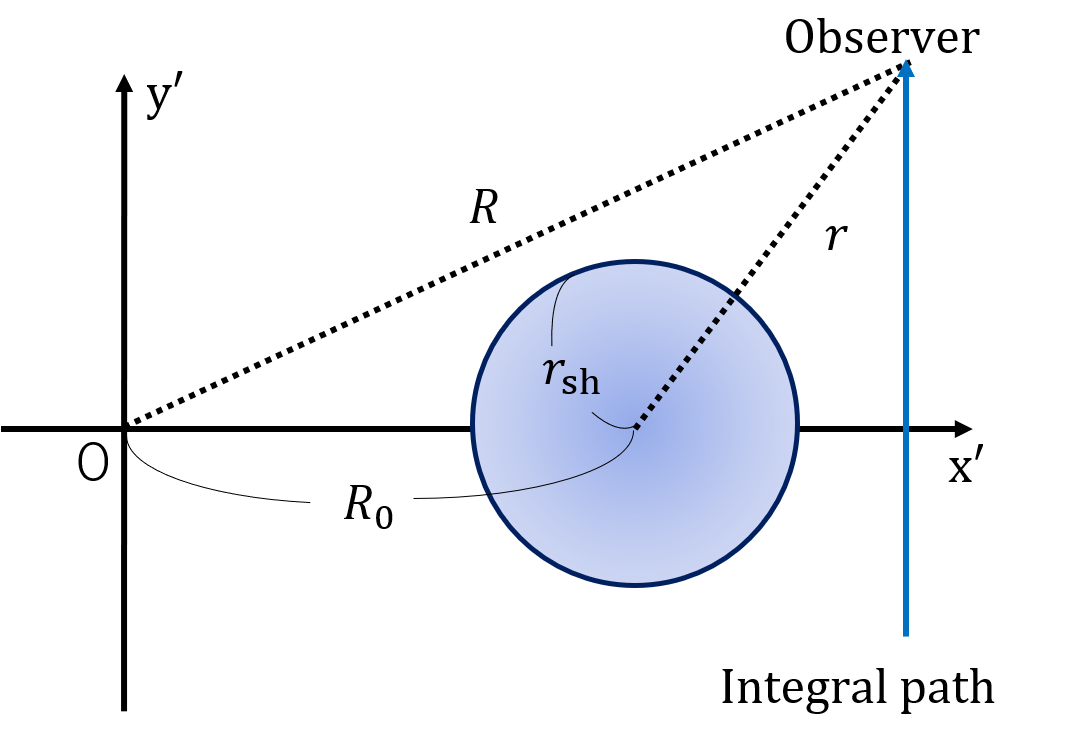}
\caption{Geometry cut along the LOS for {\tt ModBow}. }
    \label{fig2}%
\end{figure}
%%------------------------------------------------------------%%

In the case of a forward shock of the AGN jet, the shock curvature is significantly lesser than that of the host cluster.  
Because the density slope $\alpha_2$ must obey the ICM profile, we set the origin as the center of the cluster.
Thus, we slightly modify the {\tt BknPow} model, referring to it as the {\tt ModBkn} model, to optimize the jet-ICM system as follows:
\begin{equation}
    n_{\rm e}(r) = \begin{cases}
        n_0 \qty(\frac{r}{r_{\rm sh}})^{-\alpha_1} & \qty(r < r_{\rm sh}) \\
        \frac{n_0}{C} \qty( \frac{R}{R_0 + r_{\rm sh}} )^{- \alpha_2} & (r > r_{\rm sh}),   
    \end{cases}
\end{equation}
where $r$, $R$, and $R_0$ are projected radius vector from the spherical center, projected radius vector from AGN, and projected shock distance to the AGN, respectively
(see Figure \ref{fig2}).
This model is very similar to the model proposed in \citet{2013ApJ...764...82B}, where the authors adopt the model for a better fit at the shock fronts, which are less curved than the cluster.

As we already mentioned in section \ref{sec:method-xray}, studies of X-ray observations have been assumed that the X-ray brightness is independent of the electron temperature (see equation \eqref{eq:SB}).
Thus, under this assumption, the shock compression factor $C$ can be obtained by fitting the surface brightness.
In addition, the discussion of the absolute value of the X-ray surface is nonsense to determine the shock compression factor.
Since our numerical model is aimed at comparing with the {\it Chandra} broadband (0.5 -- 7.5 keV) observational data of the Cygnus A, we adopt same approximation (equation\eqref{eq:SB}) for the fiducial calculation.
However, it is necessary to investigate the influence of temperature variations in the ICM across the shock on the measurements.
We therefore confirm validity of this approximation in section \ref{sec:appendA}.

The shock Mach number is determined by the Rankine-Hugoniot jump conditions as 
\begin{equation}
    C = \frac{n_2}{n_1} = \frac{\gamma_{\rm gas}+1}{\gamma_{\rm gas}-1+2/\mathcal{M}^2},
\end{equation}
where $n_1$ and $n_2$ are the pre- and post-shock density, respectively.
The adiabatic index $\gamma_{\rm gas}$ is. Set to $5/3$ in this study.
Fitting was performed with the non-linear least-square minimization Python package 
lmfit \citep{matt_newville_2021_5570790}.
The shock Mach number can be measured from the spectroscopic-like temperature as
\begin{equation}
    \label{eq:T-jump}
    \frac{T_2}{T_1} = \frac{[(\gamma-1) \mathcal{M}^2 + 2][2 \gamma \mathcal{M}^2 - (\gamma -1)]}{(\gamma+1)^2 \mathcal{M}^2 }
\end{equation}
where $T_1$ and $T_2$ are the pre- and post-shock temperature, respectively.
The measured Mach numbers from the density and temperature jumps can be different due to several factors, such as the projection effect and temperature non-equilibrium between protons and electrons.
Some observations compared the results of two measurements \citep{2017A&A...600A.100A}.
Although the above discussion is based on the theory for hydrodynamic shock, the influence of the magnetic field is expected to be very small for the forward shock.
The observations indicate that the plasma beta of the ICM is very high \citep{2004IJMPD..13.1549G}.
We further confirmed that the plasma beta of the post-shock region is significantly higher than 100 in our MHD simulation.

To extract the radial profile of the (averaged) X-ray surface brightness and the spectroscopic-like temperature, the region is taken to be a partial annulus, whose central angle is 120 degrees, around an X-ray discontinuity by hand. 
The radial profile is computed in increments of 1 kpc in the annulus.
We try to set the curvature of the X-ray discontinuity and the partial annual to be almost the same. 
These procedures have made use of SAOImage DS9, developed by the Smithsonian Astrophysical Observatory.

\section{Results of MHD simulation}
\label{sec:result}
In this study, we focus on the X-ray property of the forward shock.
From our MHD simulation performed in Paper I, we first summarize the details of the thermodynamical evolution of the shocked-ICM, which emits enhanced thermal X-ray emission. 
In particular, temperature non-equilibrium between protons and electrons is important for thermal emission.
Then, we report the time evolution of the Mach number of the forward shock to compare the measured Mach number from the mock X-ray observation in Section \ref{sec:mockX}. 

\subsection{Thermodynamics of shocked-ICM}
\label{sec:method-ICM}
\begin{figure*}
\centering
\includegraphics[width=0.9\textwidth]{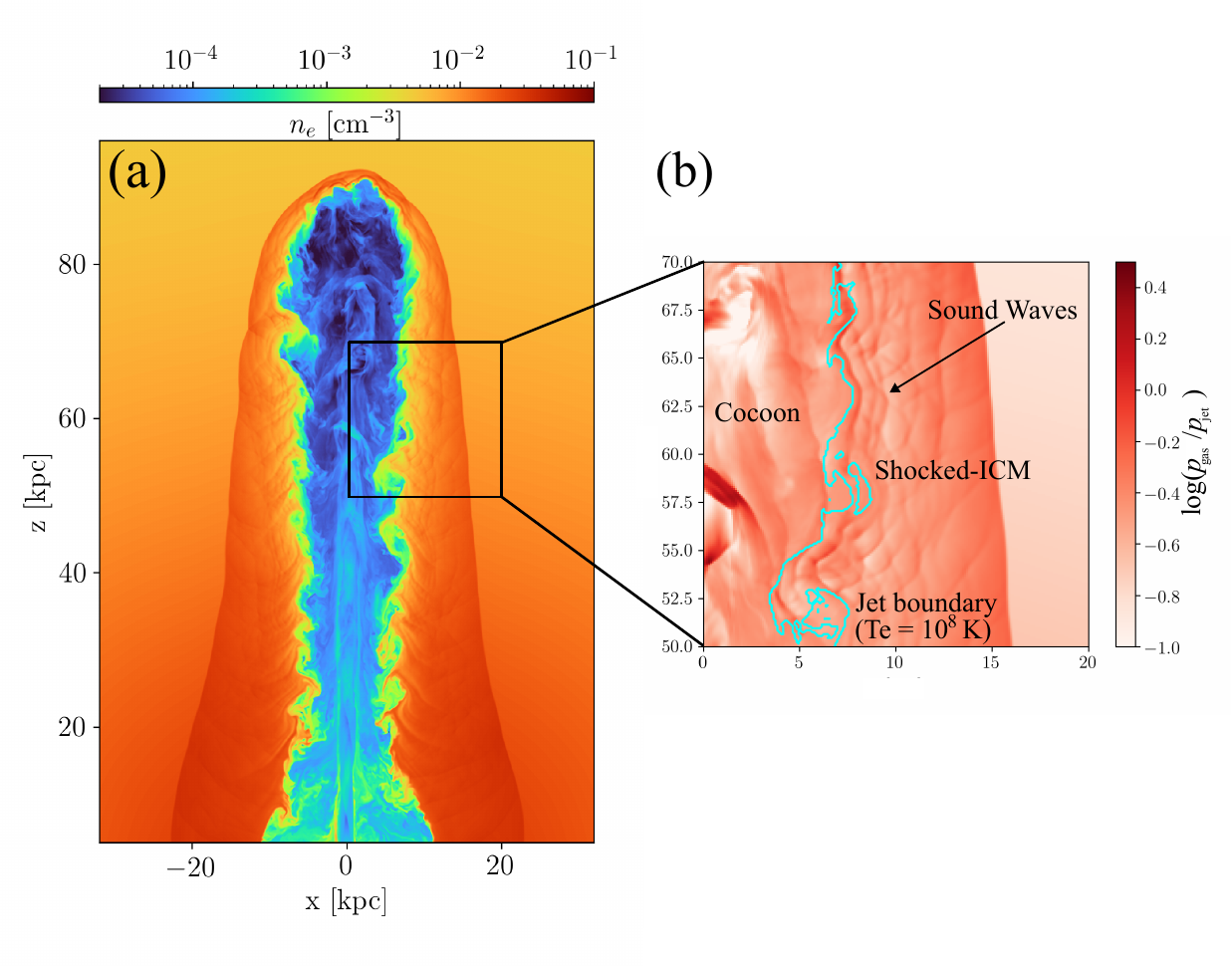}
\caption{ The structure of the simulated jet.
(a) Slice of number density distribution in the x-z plane ($y = 0$ kpc) at $t = 9.94$ Myr. (b) Slice of gas pressure distribution in x-z plane ($y = 0$ kpc). Supersonic turbulent motions of the cocoon, which is shocked jet gas, drive sound waves into shocked-ICM.}
    \label{fig3}%
\end{figure*}

We elucidate the time evolution of the shocked-ICM, especially for thermodynamic balance.
First, we show the number density distribution at $t = 9.94$ Myr in Figure \ref{fig3}a.
The forward shock drives into the ICM and has an elliptical shape, which is the usual image obtained in AGN jet simulations and is consistent with FR II radio sources.
Electrons and protons are heated by the forward shock.
Notably, protons are first hotter than electrons in the shocked-ICM, as the shocks primarily heat protons in our simulations.
Subsequently, electrons and protons approach temperature equilibrium through Coulomb collisions (see more details in next paragraph).  
The shocked-ICM still exhibits a different temperature between electrons and protons in the region of $ z > 40$ kpc when jets reach $z\sim 90$ kpc.
The sound wave is another important heating source of protons for the shocked-ICM.
The origin of sound wave production is the supersonic turbulence motion of cocoon plasma (see Figure \ref{fig3}b).
Because plasma-$\beta$ is very high in the shocked-ICM, the fraction of the electron heating is almost zero.
Thus, sound waves selectively heat protons.
A more detailed analysis of the dissipation of sound waves driven by the AGN jet is given in previous studies \citep{2005ApJ...630L...1F,2019ApJ...886...78B,2022arXiv220105298W}.

\begin{figure}
\centering
\includegraphics[width=1.0\columnwidth]{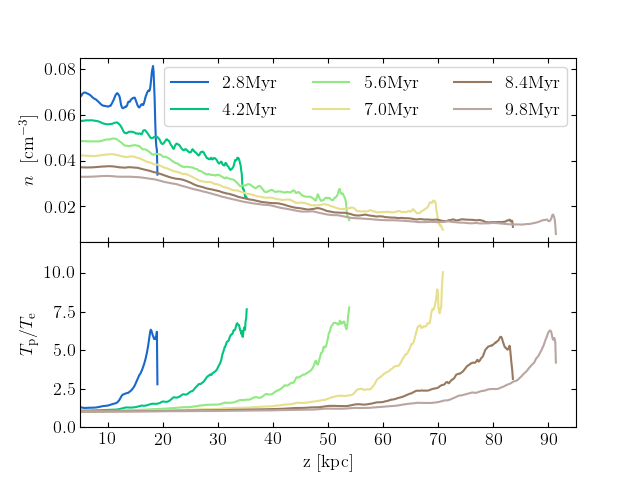}
\caption{
Averaged density profile of shocked-ICM along z-axis ({\bf top}) and the averaged ratio of proton to electron temperature ({\bf bottom}) as $t= 2.8$, 4.2, 5.6, 8.4, and 9.8 Myr.}
    \label{fig4}%
\end{figure}

To analyze further details for ICM thermodynamics, we define the shocked-ICM as the grids where $T_{\rm e} < 10^8$ K and $n(t=t') - n(t=0) > 0.05n_0$, where $t'$ is the current time.
We distinguish between the cocoon and the ICM by using the first criterion.
Also, from the second criterion, the ICM region is further divided into the shocked-ICM region and the non-perturbed ICM region.  
Figure \ref{fig4} shows the averaged density of the shocked-ICM (top) and the averaged ratio of proton to electron temperature of the shocked-ICM (bottom) along the z-axis for model B at $t = 2.8,~4.2,~5.6,~8.4,$ and $9.8$ Myr, respectively.
Herein, the averaged quantities $q$ of the shocked-ICM along the z-axis is calculated in the form,
\begin{equation}
  <q(z)> = \frac{ \int \int q dx dy}{ \int \int dx dy}
\end{equation}
${\rm for}~~T_{\rm e} < 10^8~{\rm K}$ and $n(t=t') - n(t=0) > 0.05 n_0$.
The initial density profile is the $\beta$-model, and thus the averaged density profiles of the shocked-ICM along the z-axis have a lower value, as they are father from the core (see in the top panel of figure \ref{fig4}).
The density has the highest value at the tips of the bow shock.
In particular, the shock compression is effective in the early phase.
The averaged density of the shocked-ICM likewise decreases with time due to adiabatic expansion.
Electrons and protons do not have same temperature in the shocked-ICM during simulation time (see the bottom panel of Figure \ref{fig4}).
Around the tips of the forward shock, the temperature ratio between protons and electrons is about six for all plots.
Meanwhile, electrons and protons reached thermal equilibrium, $T_{\rm e}=T_{\rm p}$, in the area close to the core due to Coulomb coupling.
Notably, a timescale of proton-electron temperature equilibrium is less than 5 Myr (see Equation \ref{eq:teq}).  

\begin{figure}
\centering
\includegraphics[width=0.95\columnwidth]{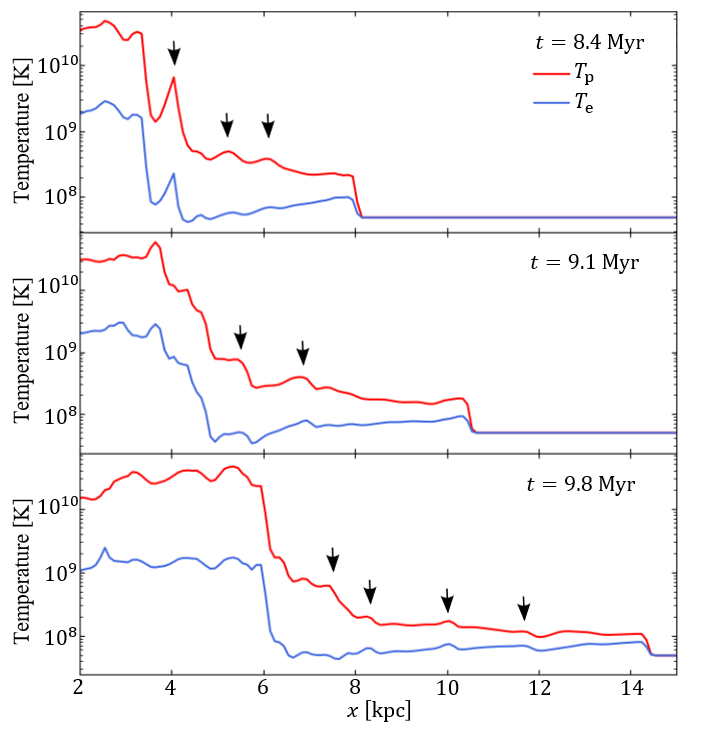}
\caption{Radial profile for electron (blue) and proton (red) temperature for model B at t =$~8.4$ ({\bf top}),$~9.1$ ({\bf middle}), and $~9.8$ ({\bf bottom}) Myr, respectively. All panels plotted along x-axis at $z = 60$ kpc and $y = 0$ kpc. Black arrows indicate the location of the sound wave.
}
    \label{fig5}%
\end{figure}

In Figure \ref{fig5}, we plot radial profile for electron (blue) and proton (red) temperature at t =$~8.4$ (top), $~9.1$ (middle) $~9.8$ (bottom) Myr, respectively.
All panels plot temperatures along x-axis at $z = 60$ kpc and $y = 0$ kpc.
As mentioned above, protons receive a large amount of the dissipation energy of the shock and sound waves, rather than electrons.
Strong sound waves propagate in the shocked-ICM from the cocoon to the forward shock, as shown in the top panel of Figure \ref{fig5} at $z= 4$, 5, and 6 kpc.
Thus, the proton temperature at the shocked-ICM decreases in the radial direction.
Meanwhile, electrons cannot receive heating energy through the sound wave.
Therefore, the electron temperature of the shocked-ICM increases in the $r$ direction, and it is maximum at the front of the shock.
Note that the sudden increase in electron temperature at the same location of the sound wave is due to adiabatic compression, not the shock heating.

\subsection{Evolution of forward shock}
\label{sec:method-shock}
\begin{figure}
\centering
\includegraphics[width=1.0\columnwidth]{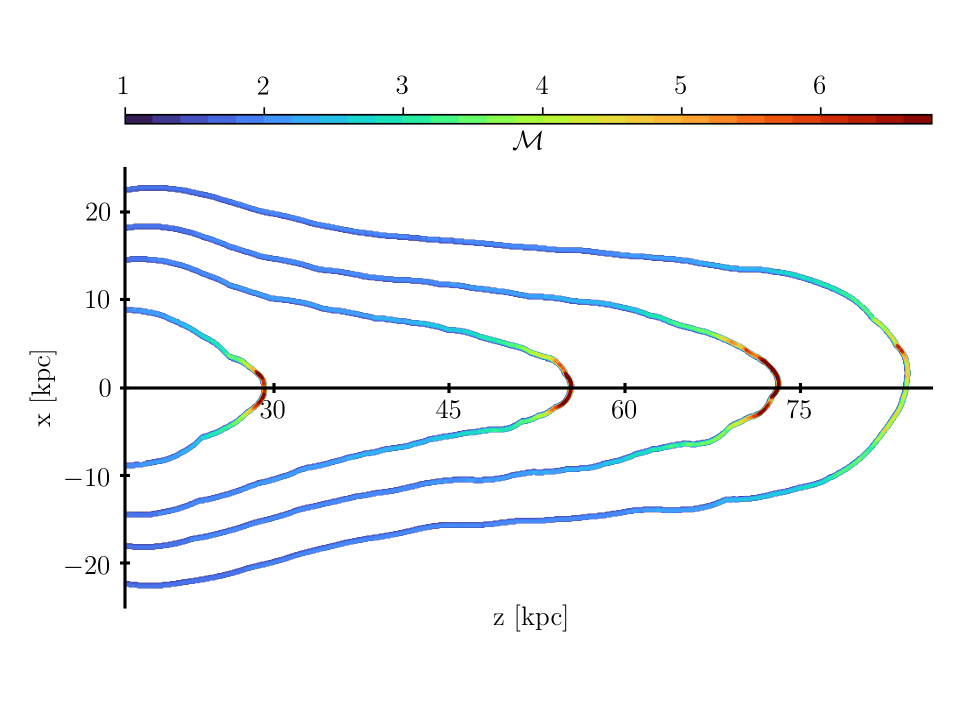}
\caption{Counters of the forward shock Mach number in the x-z plane at $t =$ 2.80, 5.60, 7.84, and 9.94 Myr from left to right.}
    \label{fig6}%
\end{figure}

We show the time evolution of the forward shock Mach number at xz-plane in Figure \ref{fig6}.
The shock-finding algorithm is adopted to determine the shock Mach number, as described in \citet{2003ApJ...593..599R} and \cite{2015MNRAS.446.3992S}.
The Mach number of the forward shock around the jet head is 6--7 for $t < 8$ Myr, as that of the injected beam is 6.3 for our model.
The high Mach number region is only a small fraction of the forward shock.
At the side of the forward shock, the Mach number has smaller value, which is about two.
After $t > 8$ Myr, the jet deteriorates due to suffered kink instability (see details in Paper I).
Thus, the shape of the forward shock becomes wider, and the Mach number around the jet head is slightly reduced.

To check the consistency, we evaluate the Mach number from the jet head velocity $v_{\rm head}$.
The ICM temperature is 5 keV homogeneously, and the Mach number is simply described as ${\mathcal M} = v_{\rm head}/c_{\rm s}$.
During $t = 4 - 8$ Myr, the jet head velocity $v_{\rm head}$ is ~0.025c--0.03c from our simulation data.
We derived the Mach number of 6.5--7.8.
After $t > 8$ Myr, the jet head velocity is reduced until 0.02c, which leads to ${\mathcal M} \sim 5.2$.
There are several tension points, but we obtain a roughly consistent result with that estimated from the shock jump condition, which is observed in figure \ref{fig6}. 

%%------------------------------------------------------------%%

%%------------------------------------------------------------%%

%%------------------------------------------------------------%%
%%------------------------------------------------------------%%

\section{Result of mock X-ray observation}
\label{sec:mockX}
The main focus of this section is to understand the impact of the projection and two-temperature effects on the measured Mach number.
We present the result of mock X-ray observation of our MHD model with different viewing angles, $\theta =$ 30$^{\circ}$, 45$^{\circ}$, 52$^{\circ}$ 60$^{\circ}$, 75$^{\circ}$, and 90$^{\circ}$, where $\theta$ is the angle between the LOS and the jet propagating direction.
The observed time is set $t = 9.94$ Myr because the length of the radio lobe is consistent with the apparent size of Cygnus A.
In this time, the total outburst energy of our model is $\sim 1.6 \times 10^{60}$ erg.  
We evaluate the Mach number in two different ways: by measuring the shock compression ratio from the fitting of X-ray surface brightness profile, and by  measuring the temperature jump from the spectroscope-like temperature.

\subsection{X-ray Surface brightness}
\label{sec:mockX-sb}
\begin{figure*}
\centering
\includegraphics[width=0.95\textwidth]{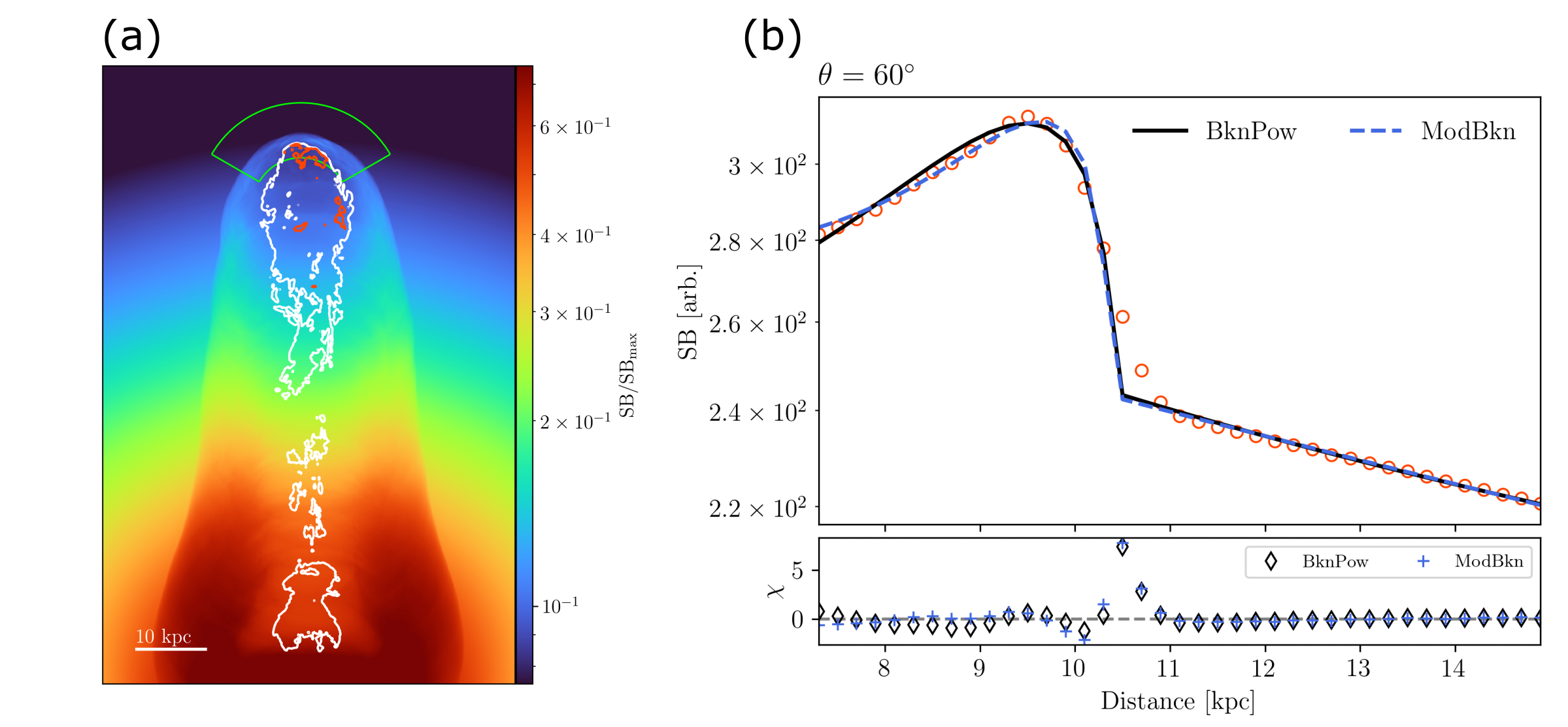}
\caption{Demonstration of mock X-ray observation. (a) Simulated X-ray surface brightness image of a AGN jet at a 60 degree viewing angle with overlaid radio intensity (white and red) contours.
Red and white contours are a 5 \% and 30 \% of maximum radio intensity, respectively.
To simplify, we assume that the radio emissivity $\propto u_{\rm e}B^2$.
The green annular sector showing the shock region to measure the surface brightness profile has been extracted.
(b) Surface brightness profile (red circle) across the forward shock and best fit results using {\tt BknPow} (black solid line) and {\tt ModBkn} models (Blue dotted line) for the sector in panel a. 
The {\tt BknPow} and {\tt ModBkn} models fit a jump with shock compression ratios $C = 2.28 \pm 0.06$ and $3.37 \pm 0.09$, respectively.
}    \label{fig7}%
\end{figure*}

In Figure \ref{fig7}a, we show simulated X-ray surface brightness image of a AGN jet at a 60 degree viewing angle with overlaid radio intensity contours (white and red). 
Adopting a simplified assumption, the radio intensity is computed to integrate a product of the thermal electron energy density and magnetic energy density along the LOS.
The forward shock compresses thermal gas, which is consequently clearly visible in X-ray map. 
Furthermore, we observe the X-ray cavity, i.e., the surface brightness is depressed spatially corresponding to projected radio lobe, as the jet plasma has low density.
Non-thermal X-ray emission, which originates from non-thermal relativistic electrons, is ignored.
Thus, the X-ray jet and hotspot are not present on this map. 
When the MHD jet remains collimated and maintains supersonic velocity at the observed time, the reverse shock survives at the jet heads (see Paper I for details).
Hence, we can observe the radio hotspot around the jet heads, because the jet kinetic energy is converted into the thermal electron and magnetic energy at the reverse shock. 

The model fitting results, which yield the shock compression parameter of the surface brightness profile across the forward shock are shown in Figure \ref{fig7}b.
The {\tt BknPow} and {\tt ModBkn} models fit the jump with shock compression ratios $C = 2.28 \pm 0.06$ and $3.24 \pm 0.08$, respectively.
The simulated X-ray discontinuity is more diffusive than both best fits.
This is due to the smearing effect over the complex structure at the edge of the shock in the sector, as the shape of the forward shock is not a perfect arc.

\begin{figure}
\centering
\includegraphics[width=1.0\columnwidth]{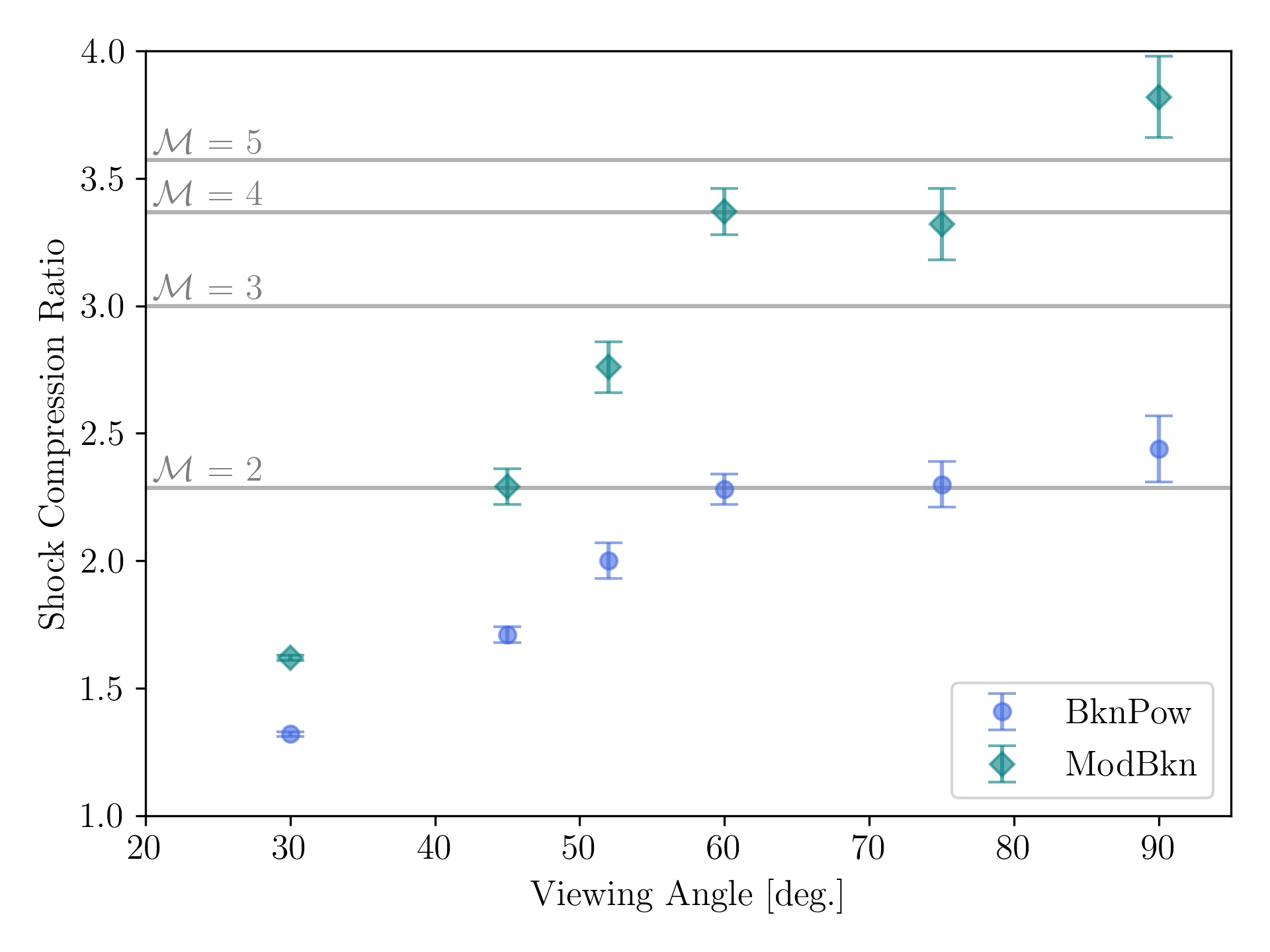}
\caption{Shock compression ratio estimated from the X-ray surface density profile as function of the viewing angle. Blue circles and green diamonds show the result obtained using {\tt BknPow} and {\tt ModBkn}, respectively. The horizontal gray lines mark the shock compression ratios corresponding the shock Mach numbers ($\mathcal M = 2, 3, 4,$ and $5$).}
    \label{fig8}%
\end{figure}

We explore the dependence of the measured Mach number on the viewing angle and show the results in Figure \ref{fig8}.
Fitting parameters are listed in Table \ref{tab:simfit}.
Herein, we mention once more that the actual shock Mach number of the simulated jet around the jet head is about five at the observed time (Figure \ref{fig6}).
At a 90 degree viewing angle, we measure the Mach number correctly using the {\tt ModBkn} model.
However, the result of the {\tt BknPow} model strongly underestimates the measured shock compression ratio compared with that of the {\tt ModBkn} model.
This tendency is observed at all viewing angles.
For both models, observed shock compression ratios strongly depend on the viewing angles, in a proportional manner.
The X-ray forward shock at a low viewing angle spatially corresponds to the side of the forward shock, not around the jet head.
From the head to the side, the Mach numbers become small (Figure \ref{fig6}).
Hence, we must focus on the effect of the viewing angle to measure the Mach number.

\begin{figure*}
\centering
\includegraphics[width=1.0\textwidth]{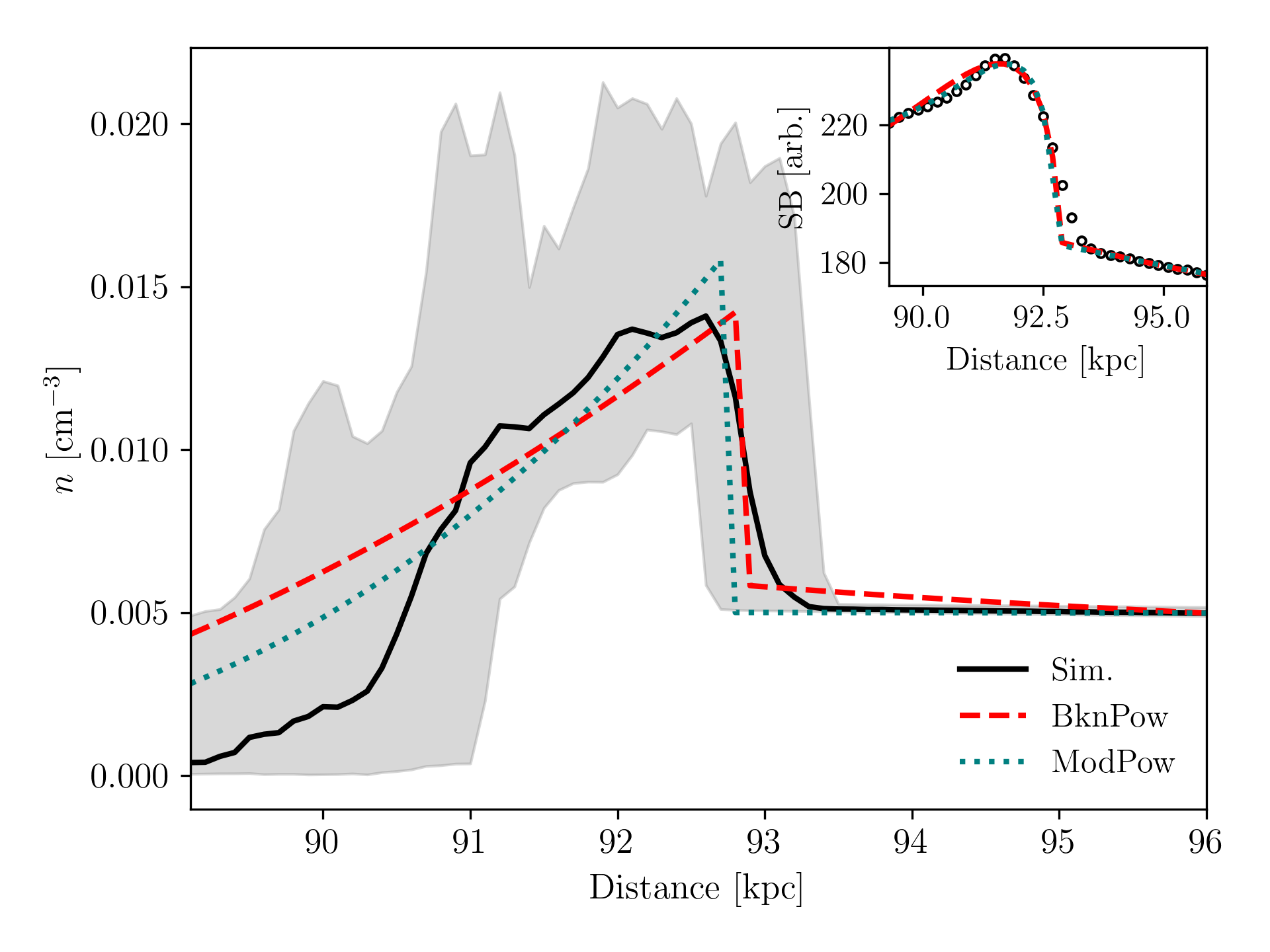}
\caption{Radial profile of thermal density. 
Red dashed and green dotted lines correspond to the best fitting {\tt BknPow} model and {\tt ModPow} model at a 90 degrees viewing angle, respectively.
A sector to measure the surface brightness profiles is shown in Figure \ref{fig_a1} at a 90 degree viewing angle.
We measure the thermal density profile from the simulation result at the x-z plane (y = 0 kpc) in the sector.
The shaded region shows the maximum and minimum thermal density, and the black solid line represents the average value.
The {\tt BknPow} and {\tt ModBkn} models fit the jumps with a shock compression ratio $C = 2.44\pm0.13$ and $3.82\pm0.16$, respectively.
The surface brightness profile (black circles), the best fits of {\tt BknPow} (red dashed line) and {\tt ModBkn} (green dotted line) are shown in the inset.
}
    \label{fig9}%
\end{figure*}

To explain the model dependence for the shock compression ratio, we show the radial profile of the thermal density, derived by the simulation result and both best fits, in Figure \ref{fig9}.
We measure the thermal density profile from the simulation result at the x-z plane (y = 0 kpc) in the sector of Figure \ref{fig_a1}.
As observed in the inset, the surface brightness profile of both best fits are very similar.
However, the derived thermal density profile is different.
Because the spherical center of the {\tt BknPow} model does not spatially correspond to the AGN, the indices of ambient profile $\alpha_2$ for the {\tt BknPow} model are clearly erroneous. 
Here, we mention that the thermal density profile obeys the $n(r) \to r^{-1.5}$ at large distances.
Consequently, the {\tt BknPow} model underestimates the density jump parameter $C$ and the shock Mach number.

%-----------------  table ------------------------------
  \begin{table*}[t]
      \caption{Shock parameters of simulation data. These parameters are described in Section \ref{sec:method-fit}  }
         \label{tab:simfit}
         \begin{tabular}{c|cccc|ccccc} \hline 
         & \multicolumn{4}{c|}{{\tt BknPow}} & \multicolumn{5}{c}{{\tt ModBkn}}  \\ \hline 
  $\theta$ & $C$ & $\alpha_1$ & $\alpha_2$ & $r_{\rm sh}$ & $C$ & $\alpha_1$ & $\alpha_2$ & $r_{\rm sh}$  & $R_0$    \\ \hline\hline
$30^{\circ}$  & $1.32\pm0.01$ & $-0.22\pm0.01$ & $0.68\pm0.04$ & $11.0 \pm 0.03$ 
              & $1.62\pm0.01$ & $-0.71\pm0.01$ & $1.32\pm0.01$ & $10.9 \pm 0.01$ & 40.0\\
$45^{\circ}$  & $1.71\pm0.03$ & $-1.61\pm0.06$ & $0.63\pm0.11$ & $11.2 \pm 0.02$ 
              & $2.29\pm0.07$ & $-2.25\pm0.11$ & $1.32\pm0.06$ & $11.3 \pm 0.00$ & 55.9\\
$52^{\circ}$  & $2.00\pm0.07$ & $-2.29\pm0.10$ & $0.58\pm0.02$ & $10.9 \pm 0.01$ 
              & $2.76\pm0.10$ & $-3.36\pm0.19$ & $1.33\pm0.08$ & $10.9 \pm 0.01$ & 62.9\\
$60^{\circ}$  & $2.28\pm0.06$ & $-3.49\pm0.14$ & $0.57\pm0.01$ & $10.4 \pm 0.03$ 
              & $3.37\pm0.09$ & $-5.42\pm0.27$ & $1.39\pm0.05$ & $10.4 \pm 0.05$ & 69.5\\
$75^{\circ}$  & $2.30\pm0.09$ & $-2.56\pm0.19$ & $0.54\pm0.02$ & $9.44 \pm 0.02$ 
              & $3.32\pm0.14$ & $-3.31\pm0.29$ & $1.38\pm0.10$ & $9.44 \pm 0.02$ & 79.4\\              
$90^{\circ}$  & $2.44\pm0.13$ & $-1.85\pm0.07$ & $0.47\pm0.02$ & $7.71 \pm 0.04$  
              & $3.82\pm0.16$ & $-2.84\pm0.12$ & $1.20\pm0.10$ & $7.75 \pm 0.03$ & 85.0\\\hline
         
         \end{tabular}
  \end{table*}
 %---------------------------------------------------
%%------------------------------------------------------------%%

%%------------------------------------------------------------%%

\subsection{Spectroscopic-like temperature}
\label{sec:mockX-tspec}
From X-ray observations, we can measure the shock Mach number using the spectroscopic (observed) temperature jump in Equation \eqref{eq:T-jump}, not only using the density jump estimated form the surface brightness profile.
The shocked-ICM around the jet head is in a two-temperature state, where the electron temperature is lower than proton temperature at the observed time (see section \ref{sec:method}).
Hence, we calculate the spectroscopic-like temperature in two ways: by using the electron temperature in Equation \eqref{eq:tspec} in a straightforward manner, and by using the gas temperature, $T_{\rm gas} = 0.5(T_{\rm p} + T_{\rm e})$, instead of the electron temperature. This case is same that the one assuming one-temperature plasma, i.e., temperature equilibrium between the electron and proton.
Here, we mention once more that in our simulation, the electron can receive 5 \% of the dissipated energy by shocks from the proton.

\begin{figure*}
\centering
\includegraphics[width=1.0\textwidth]{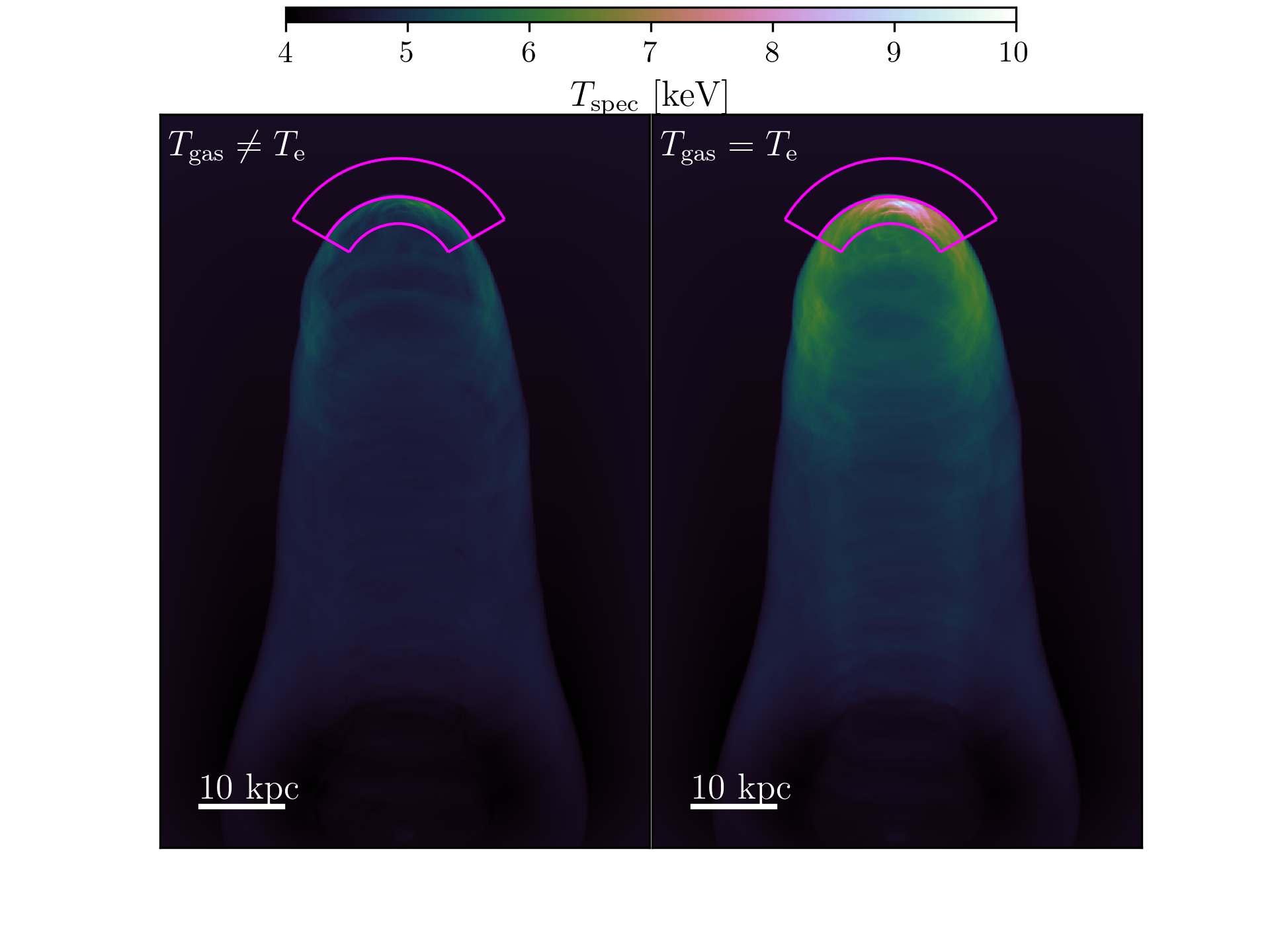}
\caption{Spectroscopic-like temperature maps in the case of two-temperature shock ({\bf left}) and one-temperature shock ({\bf right}) at a 60 degree viewing angle.
The magenta annular sector is used to measure the extracted post- and pre-shock temperatures, and the middle arc shows the position of X-ray discontinuity.
}
    \label{fig10}%
\end{figure*}

In Figure \ref{fig10}, we show spectroscopic-like temperature maps in the case of the two- and one-temperature shocks at a 60 degrees viewing angle.
The discontinuity of the spectroscopic-like temperature can be observed for both maps.
The electron temperature at the forward shock around the jet head is about $2.0 \times 10^8$ K for the two-temperature plasma case, and $1.0\times10^9$ K for the temperature equilibrium case.
These values are consistent with the shock jump condition (we describe the temperature jump condition for the two-temperature shock in the Appendix of \citet{2020MNRAS.493.5761O}).
Meanwhile, the spectroscopic-like (observed) temperature is significantly lower than the electron temperature, due to the projection effect.
Because the pass length at the shocked-region is smaller than that at the unshocked region, the foreground, and background ICM, which have a lower temperature than the shocked gas, contribute largely to the observed temperature.
Moreover, we find that the observed temperature for the temperature equilibrium case is slightly higher than that of the two-temperature case.

\begin{figure}
\centering
\includegraphics[width=1.0\columnwidth]{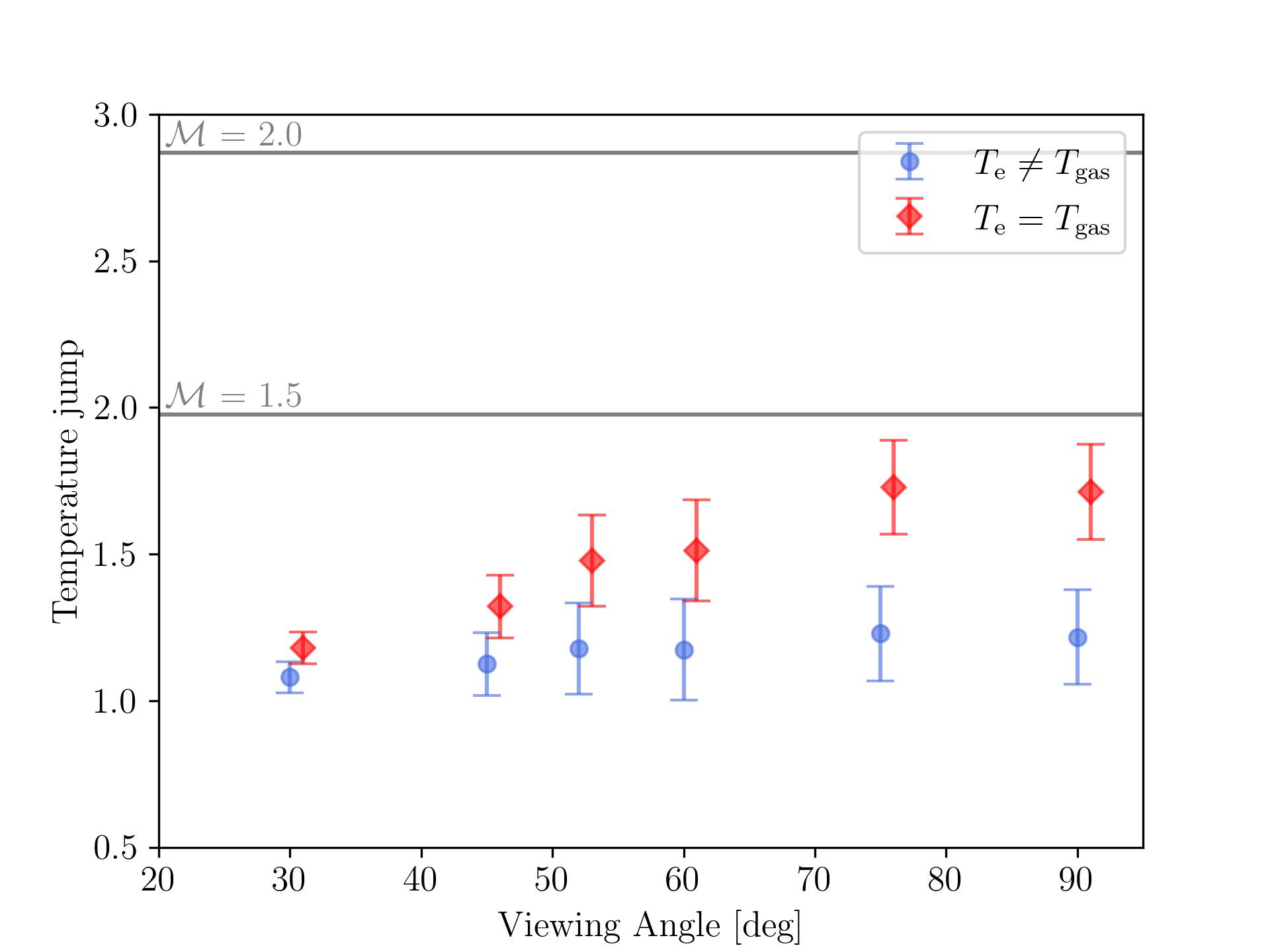}
\caption{Temperature jump ratio estimated from the spectroscopic-like temperature as a function of the viewing angle. Blue circles and red diamonds show the result in the case of two- and one-temperature shocks, respectively.  The horizontal gray lines mark the temperature jump ratios corresponding the shock Mach numbers ($\mathcal M = 1.5$, and $2$).}
    \label{fig11}%
\end{figure}

The dependence of the temperature jump on different viewing angles is plotted in Figure \ref{fig11}.
We measure the spectroscopic-like temperature at magenta the sectors inside and outside of the forward shocks, shown in Figure \ref{fig_a2}, to determine the temperature jump. 
At first glance, the Mach number measured from temperature jump is significantly underestimated for all cases.
Figure \ref{fig11} illustrates the same tendency as in the shock compression ratio (Figure \ref{fig8}).
While the plasma has a one-temperature, i.e., the efficient Coulomb collision case, the measured temperature jump is lower than a factor two, even if the viewing angle is high.
Namely, the observed shock Mach number from the temperature jump is below 1.5.
However, it is difficult to detect the temperature jump in the case temperature non-equilibrium.
The effect of the viewing angle for the observed temperature is almost same as discussed in Section \ref{sec:mockX-sb}.
Nevertheless, there is an additional factor reducing the temperature jump.
When the viewing angle is low, the projected distance from AGN to the jet head becomes short.
Consequently, the LOS passes through a high density region, which contributes largely to the observed temperature.

%%------------------------------------------------------------%%
%%------------------------------------------------------------%%

\subsection{Validity of the approximation for the thermal emission} \label{sec:appendA}

In this subsection, we discuss the effect of the electron temperature dependence of the X-ray emissivity (see also figure \ref{fig1}).
We made X-ray maps of $\theta = 90^{\circ}$ with three different energy bands (0.5 -- 3.5 keV, 3.5 -- 7 keV, and 0.5 -- 7.0 keV), and performed model fit of the X-ray surface brightness in the same forward shock region.  
The model fitting results are shown in Table \ref{tab:append_simfit}.
The ICM and shocked-ICM is about 5 keV and 10-50 keV in the case for a single-fluid case.
For the 3.5 -- 7 keV, the cooling curve monotonically increase by factor two or three between 5 keV to 50 keV.
As a results, the shock compression ratios is larger than 4, which is a maximum shock compression in the hydrodynamics.
In contrast, the electron temperature dependence of the X-ray emissivity in the temperature range we interest is weak for the 0.5 -- 3.5 keV and 0.5 -- 7.0 keV.
We can see that the compression ratios observed in these bands differ by about $\pm 0.2$, compared with the result by using the simple approximation.
Note that if we assume the two-temperature plasma, this difference become small.
This is because the electron temperature for two-temperature plasma is lower than that for single-temperature plasma.
Therefore, it is not so bad to ignore the temperature dependence of X-ray emissivity, unless we compare it to data with 3.5 -- 7.0 keV.

%-----------------  table ------------------------------
  \begin{table*}[t]
      \caption{Shock parameters with varying the observed energy bands. The inclination angle sets $\theta = 90^{\circ}$.   }
         \label{tab:append_simfit}
         \begin{tabular}{c|cccc|cccc} \hline 
         & \multicolumn{4}{c|}{{\tt BknPow}} & \multicolumn{4}{c}{{\tt ModBkn}}  \\ \hline 
  Energy band & $C$ & $\alpha_1$ & $\alpha_2$ & $r_{\rm sh}$ & $C$ & $\alpha_1$ & $\alpha_2$ & $r_{\rm sh}$     \\ \hline\hline
$\varepsilon \propto n^2_{\rm e}$  & $2.44\pm0.13$ & $-1.85\pm0.07$ & $0.47\pm0.02$ & $7.71 \pm 0.04$  
              & $3.82\pm0.16$ & $-2.84\pm0.12$ & $1.20\pm0.10$ & $7.75 \pm 0.03$ \\\hline
0.5 -- 3.5 keV& $2.27\pm0.10$ & $-1.58\pm0.06$ & $0.47\pm0.02$ & $7.94 \pm 0.05$ 
              & $3.26\pm0.11$ & $-2.55\pm0.10$ & $1.07\pm0.06$ & $7.82 \pm 0.41$ \\
3.5--7 keV    & $4.14\pm0.13$ & $-2.32\pm0.14$ & $0.35\pm0.01$ & $7.83 \pm 0.06$ 
              & $5.58\pm0.16$ & $-2.67\pm0.14$ & $0.68\pm0.02$ & $7.81 \pm 0.05$ \\
0.5--7 keV    & $2.70\pm0.08$ & $-1.84\pm0.06$ & $0.44\pm0.01$ & $7.92 \pm 0.04$ 
              & $3.82\pm0.10$ & $-2.63\pm0.10$ & $0.98\pm0.04$ & $7.82 \pm 0.04$ \\ \hline  
         \end{tabular}
  \end{table*}
 %---------------------------------------------------
%%------------------------------------------------------------%%
%%----------------------------------%%

\section{Application for forward shock of Cygnus A}
\label{sec:cygAshock}
First, we present the fitting results of the forward shock of Cygnus A.
In the previous section, we show that the shock Mach number could be determined with good accuracy using the {\it Chandra} broadband X-ray surface brightness profile because the X-ray emissivity has weak dependence with the electron temperature in this observed range.
Furthermore, the {\tt ModPow} model can evaluate the measured Mach number more accurately than the {\tt BknPow} model.
Thus, we adopt this model in the actual X-ray data of Cygnus A to investigate model dependence.
To perform the investigation, we use archival {\it Chandra} observation data in an energy band of 0.5 - 7.0 keV. 
The process was performed with the latest CALDB and utilizes the Chandra tool {\tt chandra\_repro} available in CIAO.
Then, we provide a physical interpretation of Cygnus A, achieved from the comparison of mock X-ray results and actual observations. 

\subsection{Fitting results}
\begin{figure*}
\centering
\includegraphics[width=1.0\textwidth]{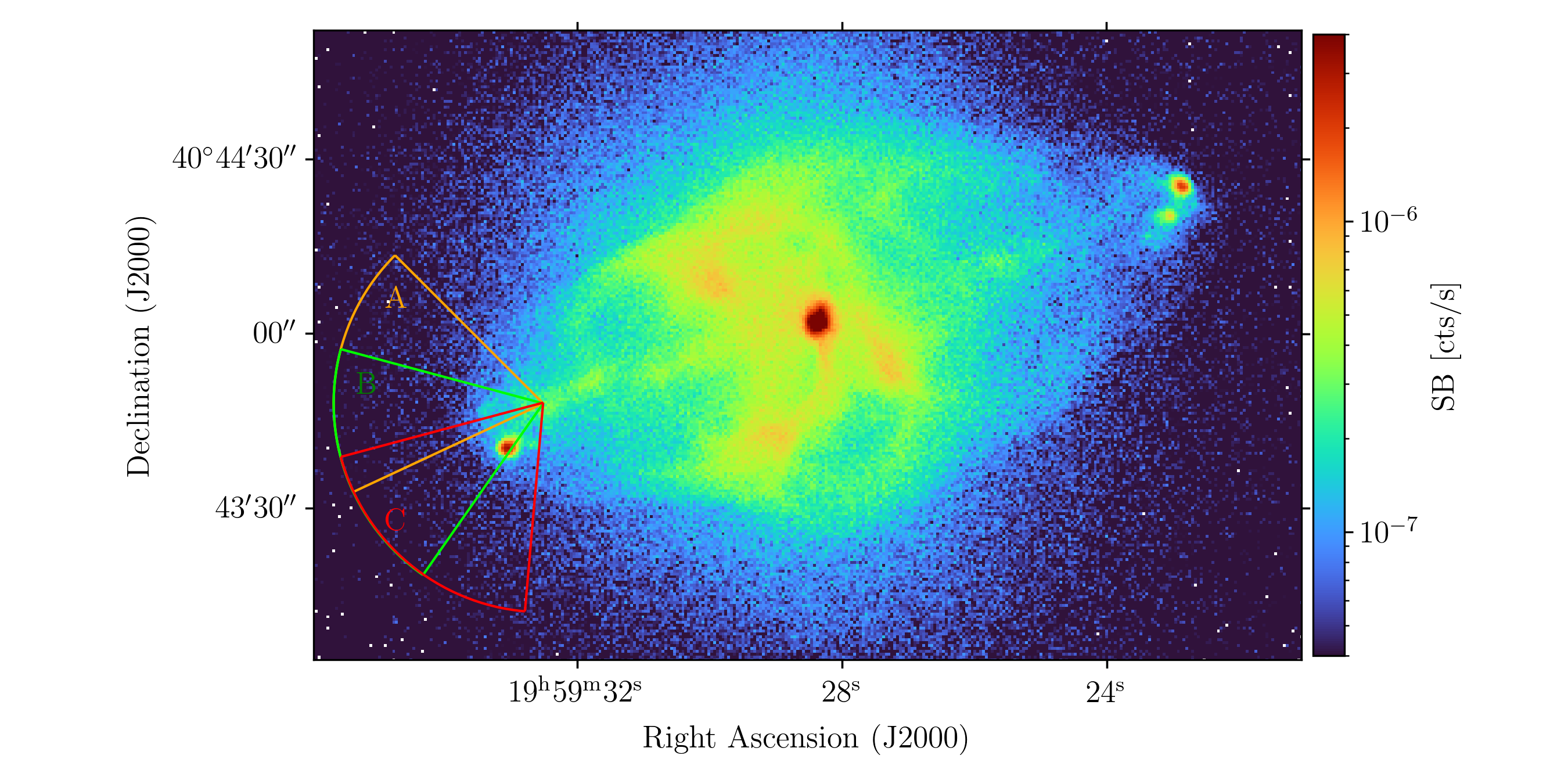}
\caption{ 0.5-7.0 keV {\it Chandra} image of Cygnus A \citep{2018ApJ...855...71S}. 
The three annular sectors showing the shock regions extracted to measure the surface brightness profiles.
}
    \label{fig12}%
\end{figure*}

Sectors to measure surface brightness profiles of the shock are marked in Figure \ref{fig12}.
The fitting results are listed in Table \ref{tab:cygA}, and shown in Figure \ref{fig_b1}.
We choose the center of sectors as the center of the X-ray jet, as follows \citet{2018ApJ...855...71S}.
For {\tt ModBkn}, the projected radius from AGN to the center of the sector $R_0$ is 0.78 arcmin.
Sector C is slightly wider than region of Sector 8 in \citet{2018ApJ...855...71S}, but it is similar.
In the Sectors B and C, the X-ray hotspot contributes largely to the surface brightness.
Hence, we eliminated the hotspot.

Shock compression ratios vary within 1.44--1.81 for {\tt BknPow}.
These values are slightly lower, but mostly consistent with the result of \citet{2018ApJ...855...71S}, which is $C = 1.82^{+0.38}_{-0.23}$.
This difference could be attributed to the various methods to choose sectors and the effect of the elimination of the hotspot.
The index of ambient profile $\alpha_2$ is approximately 0.9 in whole regions.
\citet{2002ApJ...565..195S} reported the fitting result of the radial profile of surface brightness of the whole {\it Chandra} field by an isothermal $\beta$ model and the slope parameter and angular core radius as 1.51 and 18", respectively. 
This indicates that spherical symmetry, whose origin is AGN, would be a good approximation.
The fitting values of $\alpha_2$, therefore, would not be consistent with the profile of the entire {\it Chandra} field.

In contrast, for {\tt ModBkn}, shock compression ratios vary within 1.90--2.40, and they are higher those for {\tt BknPow}, in all sectors.
The value of $\alpha_2$ is approximately 1.5, which is consistent with the profile of the whole {\it Chandra} field.
{\tt BknPow} may also provide more accurate fits, as the reduced chi square values of {\tt ModPow} are lower than those of {\tt BknPow} in all cases.
However, we cannot conclude that {\tt ModPow} is statistically better.
Finally, the measured shock distances $r_{\rm sh}$ exhibit similar values for the two models.

%%------------------------------------------------------------%%
%%------------------------------------------------------------%%

%-----------------  table ------------------------------
  \begin{table*}[t]
      \caption{Shock parameters of Cygnus A. In {\tt ModBkn}, the projected radius from AGN to the center of sector $R_0$ is 0.78 arcmin for all sectors.}
         \label{tab:cygA}
         \begin{tabular}{c|ccccc|ccccc} \hline 
         & \multicolumn{5}{c|}{{\tt BknPow}} & \multicolumn{5}{c}{{\tt ModBkn}}  \\ \hline 
         Sector & $C$ & $\alpha_1$ & $\alpha_2$ & $r_{\rm sh}$ & $\chi^2$/d.o.f & $C$ & $\alpha_1$ & $\alpha_2$ & $r_{\rm sh}$ & $\chi^2$/d.o.f    \\ \hline\hline
            A   & $1.49\pm0.05$ & $0.60\pm0.07$ & $0.89\pm0.01$ & $0.23$ & 1.10
                & $2.23\pm0.08$ & $0.64\pm0.07$ & $1.5 \pm0.02$ & $0.23$ & 1.05\\
            B   & $1.73\pm0.07$ & $0.27\pm0.11$ & $0.93\pm0.01$ & $0.22$ & 1.61
                & $2.49\pm0.11$ & $0.48\pm0.09$ & $1.6 \pm0.02$ & $0.23$ & 1.46\\
            C   & $1.54\pm0.09$ & $0.48\pm0.13$ & $0.90\pm0.01$ & $0.23$ & 1.71
                & $2.18\pm0.11$ & $0.58\pm0.10$ & $1.51\pm0.02$ & $0.25$ & 1.24\\
        A+B+C   & $1.44\pm0.04$ & $0.60\pm0.08$ & $0.90\pm0.01$ & $0.23$ & 1.44
                & $1.90\pm0.05$ & $0.82\pm0.05$ & $1.49\pm0.01$ & $0.25$ & 1.09\\\hline
         \end{tabular}
  \end{table*}

\subsection{Comparison with our jet model and physical interpretation}
Our model contributes to determine the viewing angle.
The viewing angle of Cygnus A is poorly constraint from observations, in ranging 35°--80° \citep{1995PNAS...9211371B}.
From the comparison of two results (see Table \ref{tab:simfit} and \ref{tab:cygA}, and figure \ref{fig8}), it is reasonable that the viewing angle of Cygnus A roughly ranges in 35°--55°.
Surely, part of our constraint is model dependent.
If the actual total outburst energy and power of Cygnus A is significantly higher than that of our model, the range of the viewing angle tends to be lower.

Comparing our simulations and the X-ray observation of Cygnus A, the sign of $\alpha_1$, which is the index of shock slope, is different.
The minus sign of $\alpha_1$ for Cygnus A indicates that the gas density is higher inside the forward shock than at shock front.
One explanation for this difference is that our model ignores non-thermal emission via inverse Compton scattering by relativistic electrons in the lobe.
Detailed X-ray analysis \citep{2010ApJ...714...37Y,2018MNRAS.478.4010D} shows that there is a large population of non-thermal X-ray components in the lobe and jets.
To investigate the effect of non-thermal components for the measurement of the shock Mach number, we performed same analysis using the {\it Chandra} soft-band data-set (0.5--1.2 keV), whose energy band is expected to be the small contribution of non-thermal radiation. 
Although the systematic error is large for small photon numbers, we observe that $\alpha_1 = 0.39 \pm 0,29$ in the sector B, which is slightly smaller than that observed in the energy band of 0.5 -- 7 keV.
Note that the shock compression ratio $C$ remains relatively unchanged for both energy bands.
From the simulation results, $\alpha_1$ is larger when the viewing angle is lower.
While this trend may supports that Cygnus A has a small viewing angle, the modeling of non-thermal components is needed for further discussion. 

Next, we discuss the spectroscopic temperature. 
\citet{2018ApJ...855...71S} reported the spectroscopic temperature jump of Cygnus A.
They found that the temperature jump is almost unity, which is significantly smaller than the value predicted from the measured Mach number using the shock compression.
The spectroscopic-like temperature is significant affected by the projection effect in the jet-ICM system.
These observational results are, both quantitatively and qualitatively, consistent with our two-temperature shock model, despite of a viewing angle (Figure \ref{fig11}).
Unfortunately, if the viewing angle of Cygnus A is lower than 50 degrees, as in the discussion above, it is difficult to decide whether the shocked-plasma is in the two-temperature state.

Our simulations adopt a isothermal $\beta$-model of 5 keV ICM.
However, the temperature gradient exists in the X-ray observations \citep{2018ApJ...855...71S}.
The temperature far from the AGN is hotter than that near the AGN.
The temperature gradient slightly affects the spectroscopic-like temperature map and the temperature jump at the forward shock.
As discussed in \citet{2018ApJ...855...71S}, the smearing effect over the complex structure at the edge of the shock.

\section{Summary and discussions}
\label{sec:summary}
We report the thermodynamics of shocked-ICM and the evolution of the forward shock for our two-temperature MHD simulation of AGN jets.
We performed mock X-ray observation to measure the Mach number of the forward shock from the surface brightness and spectroscopic-like temperature.
The discrepancy in the Mach number of the forward shock for the analytic and numerical model is significantly higher than that of the observation \citep{2020NewAR..8801539H}.
This study systematically investigates various effects that influence the observed Mach number.
Our results attribute this discrepancy to the projection effect, in particular, the measured shock Mach number from surface brightness profile is significantly underestimated when the viewing angle is low.

Our mock X-ray observations indicate that the measured Mach number from the surface brightness profile is very sensitive to the observed viewing angle, and monotonically increases with it.
Because the curvature of the forward shock is significantly higher than that of the cluster, we propose a new fitting model {\tt ModBkn}, instead of a traditional model {\tt BknPow}.
we demonstrate that {\tt ModPow} provides higher Mach number than one from {\tt BknPow}.
Further, the result of {\tt ModPow} is consistent with the actual shock Mach number of MHD data at the higher viewing angle.

Spectroscopic-like temperatures are calculated from our MHD data.
The projection effect significantly reduces the temperature jump, which is lower than one from the Rankine-Hugoniot condition.
Even if the electron cannot be heated instantaneously at shock front, i.e., the plasma is two-temperature at post-shock, the detection of the temperature jump is very difficult.

We estimate the shock Mach number of Cygnus A using archival {\it Chandra} observation data with {\tt BknPow} and {\tt ModPow}.
%%%%%%%%%%%
The best fit of {\tt ModPow} is shown in the three following results. Compared with that of {\tt BknPow},
(1) Shock compression ratio is high.
(2) The slope of ICM, $\alpha_2$, is consistent with the observed slope.  
(3) The reduced chi square values are small for whole regions, but we cannot conclude that {\tt ModPow} is statistically better.
%%%%%%%%%%%

Studies of X-ray observations have been assumed that the surface brightness is independent of the electron temperature to the measurement of the shock compression factor.
To check validity of approximation, we conduct mock X-ray observation with three different energy bands.
We find that the approximation is not so bad when we use the X-ray data with 0.5 -- 7.0 keV and 0.5 -- 3.5 keV.

From the mock X-ray observation results, the viewing angle of Cygnus A may range within 35°-55°degrees.
In this range, the observed Mach number and the temperature jump are consistent with our analysis of Cygnus A.
However, we note that the lower limit of a viewing angle is very rough because our X-ray maps at low viewing angles are small compared with that of actual observation. 

Our results indicate that, to construct numerical models for powerful and young radio jets like those of Cygnus A, the jet with the strong shock can be used with X-ray observations.
However, the results of this study do not directly contribute to solve the cooling flow problem, as heating by strong forward shocks is a sub-dominant source for transporting jet energy to the ICM \citep{2003NewAR..47..565H}.

In this study, we only focus on forward shock of young powerful FR II type jets. However, the {\tt ModPow} model can also be adopted for measuring the density jump of the shock and cold fronts that are more or less curved than the cluster.
Shocks of the galaxy clusters Abell 2146 \citep{2010MNRAS.406.1721R} and A 512  \citep{2013ApJ...764...82B}may be good candidates to test our fitting model.

Our model has several limitations.
Primarily, we do not treat heat conduction. 
Further, there is a high temperature gradient across a shock front.
Thus, the shock becomes diffusive, and a temperature and density precursor can form at the pre-shock region if thermal conduction works well across the shock front \citep{2020MNRAS.497.1434K}.
This would decrease the measured Mach number.
However, whether heat conduction works efficiently depends on the thermal conductivity and topology of magnetic fields.
Furthermore, if the Braginsky viscosity \citep{1965RvPP....1..205B} and turbulence motion of the ICM would be incorporated, the result would change.
The presence of cosmic rays likewise influences the dynamics.
The back-reaction from cosmic rays to the fluid modifies the shock structure, and the shock jump becomes smaller \citep{1986MNRAS.223..353D}.
In a future study, we aim to conduct MHD simulations including these omitted factors.

\begin{acknowledgements}
    We thank the anonymous referee for the useful comments that greatly improved the presentation of the paper.
    This work was supported by JSPS KAKENHI Grant Numbers JP22K14032 (T.O.) and 19K03916 (M.M.). Our numerical computations were carried out on the Cray XC50 at the Center
    for Computational Astrophysics of the National Astronomical Observatory of Japan. The computation was carried out using the computer resource by Research Institute for Information Technology, Kyushu University.
    This work was also supported in part by MEXT as a priority issue (Elucidation of the fundamental laws and evolution of the universe) to be tackled by using post-K Computer and JICFuS and by MEXT as “Program for Promoting Researches on the Supercomputer Fugaku” (Toward a unified view of the universe: from large scale structures to planets).
    SRON is supported financially by NWO, the Netherlands Organization for Scientific Research. 
\end{acknowledgements}

\bibliographystyle{aa} % style aa.bst
\bibliography{ref} % your references Yourfile.bib

\begin{appendix}

%%----------------------------------%%
\section{Best fit of simulation data}
\label{sec:appendB}
\begin{figure*}
\centering
\includegraphics[width=1.0\textwidth]{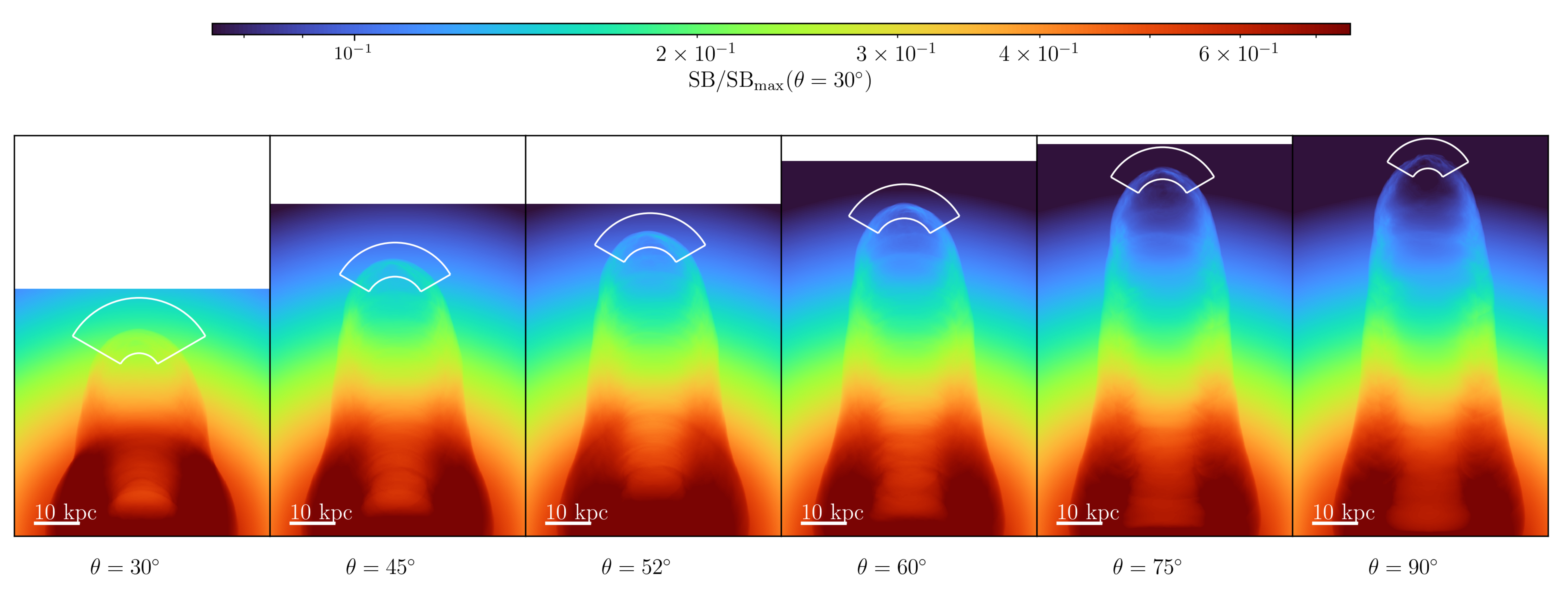}
\caption{ Simulated X-ray surface brightness maps of a AGN jet at 30, 45, 52, 60, 75, and 90 degrees viewing angles (from left to right panels), respectively.
The values for each plots are normalized by a maximum value in the case of a 30 degree viewing angle. 
The green annular sector depicts the shock regions where the surface brightness profiles have been extracted.
}
    \label{fig_a1}%
\end{figure*}

\begin{figure*}
\centering
\includegraphics[width=1.0\textwidth]{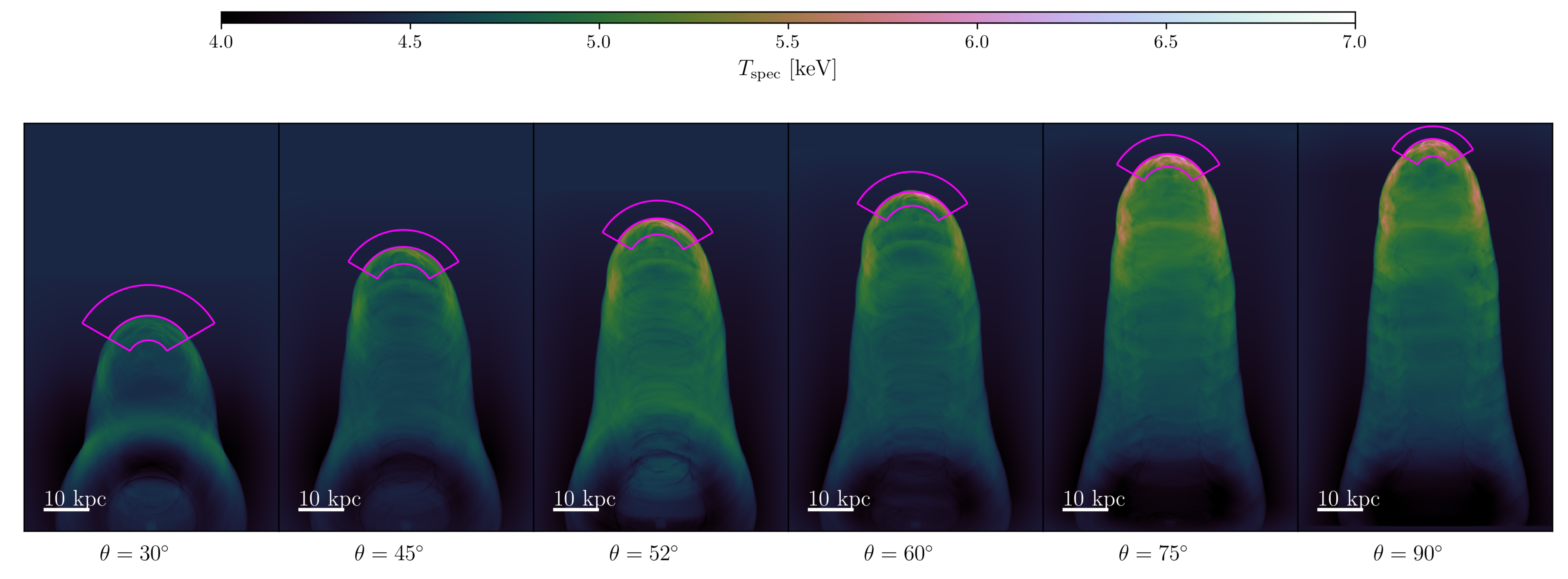}
\caption{ Spectroscopic-like temperature maps in the case of two-temperature shock at 30, 45, 52, 60, 75, and 90 degree viewing angles (from left to right panels), respectively.
The magenta annular sectors used to measure the post and pre-shock temperatures are depicted, and the middle arc shows the position of X-ray discontinuity. 
}
    \label{fig_a2}%
\end{figure*}

\begin{figure*}
\begin{tabular}{cc}
\begin{minipage}{0.48\textwidth}
\centering
\includegraphics[width=1.0\textwidth]{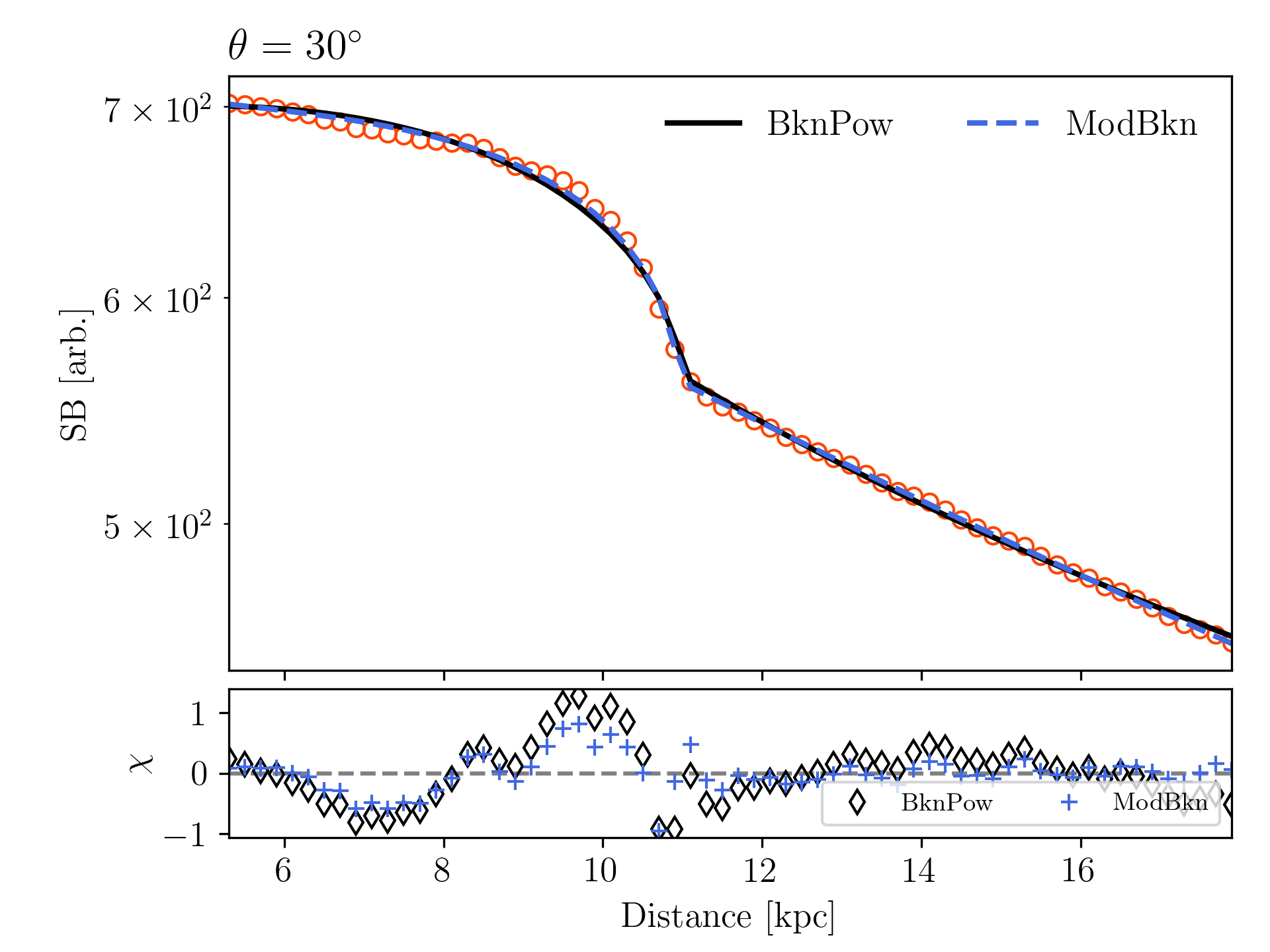}
\end{minipage} &

\begin{minipage}{0.48\textwidth}
\centering
\includegraphics[width=1.0\textwidth]{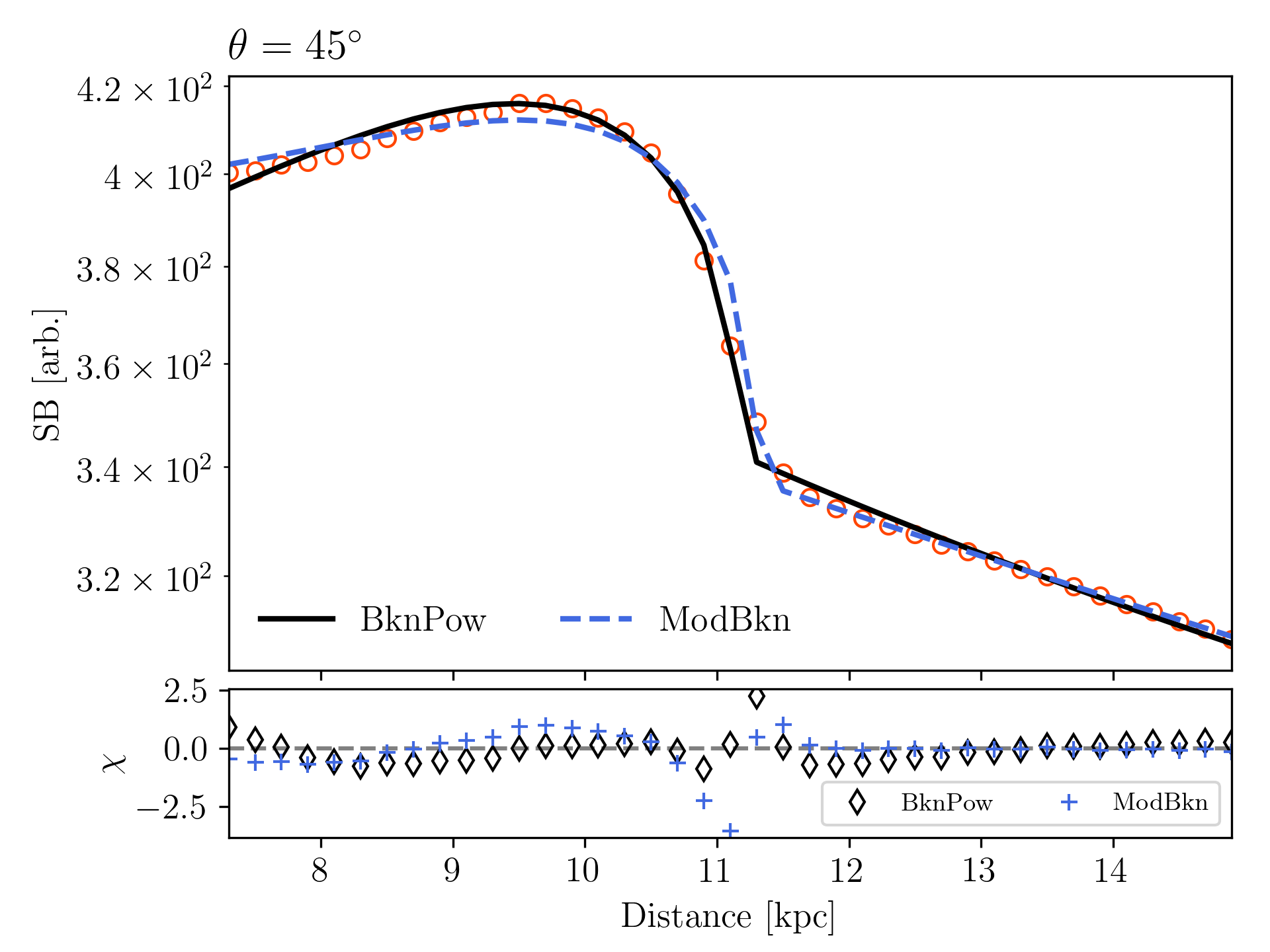}
\end{minipage}
\\
\begin{minipage}{0.48\textwidth}
\centering
\includegraphics[width=1.0\textwidth]{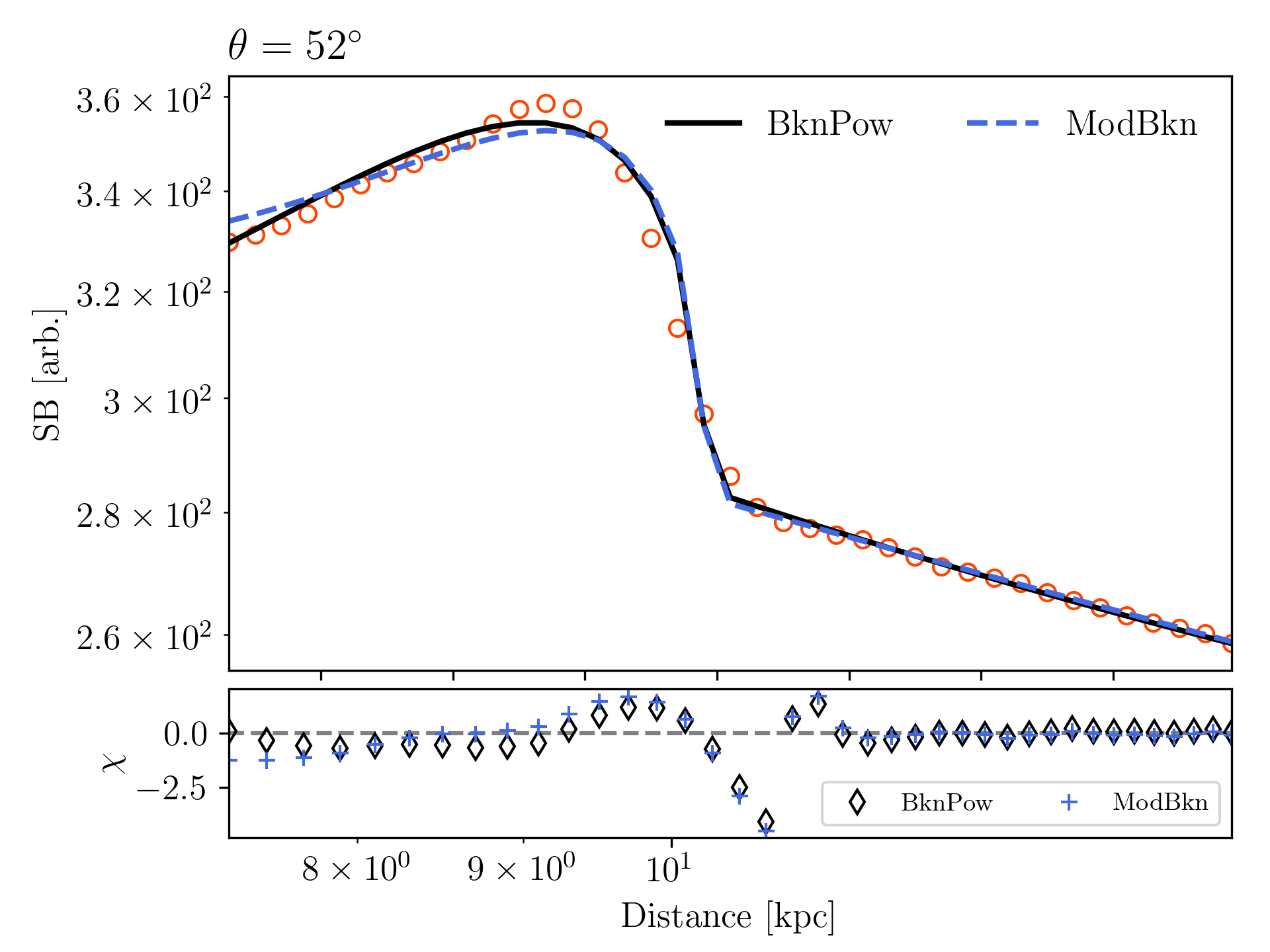}
\end{minipage} &

\begin{minipage}{0.48\textwidth}
\centering
\includegraphics[width=1.0\textwidth]{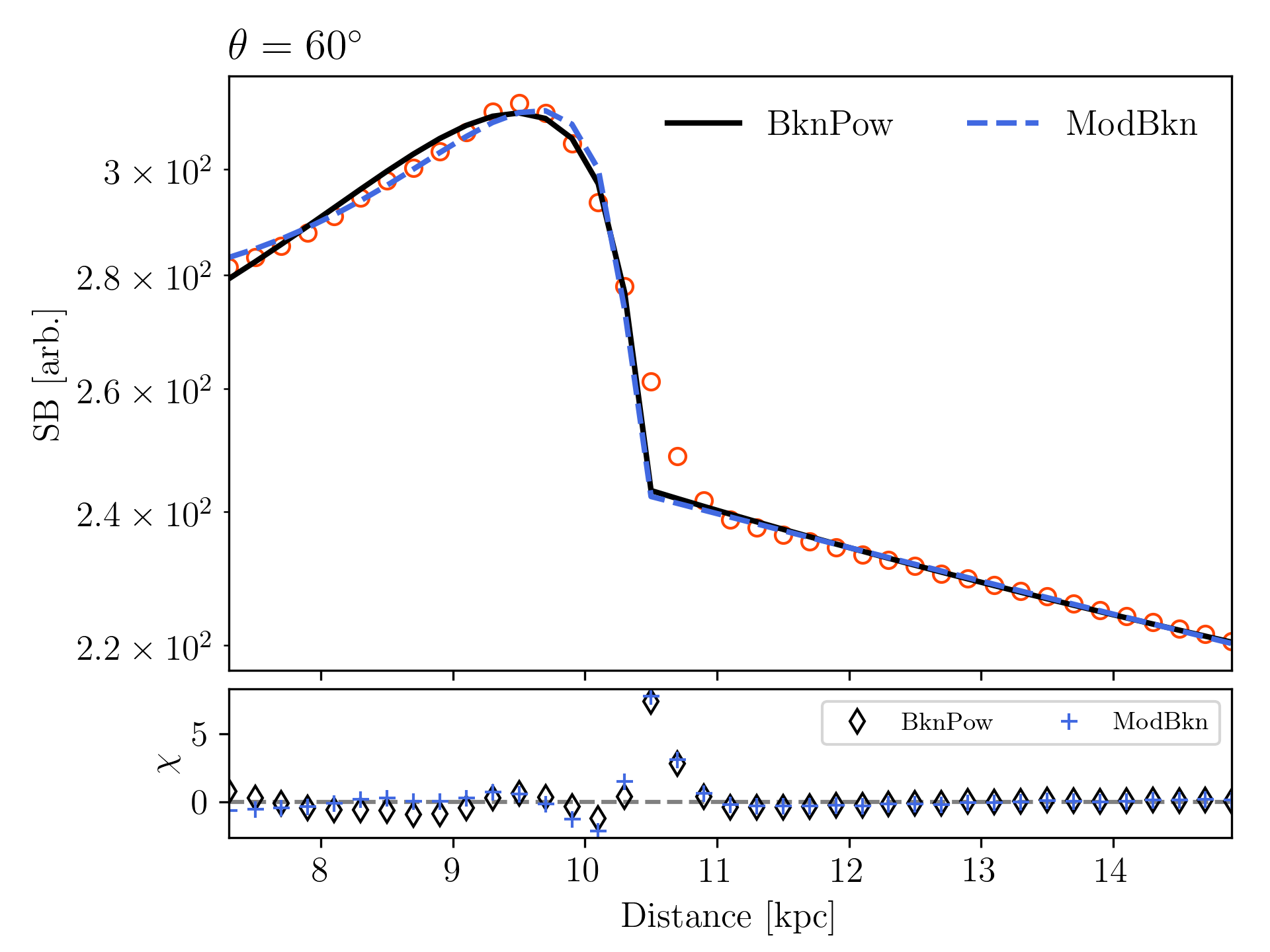}
\end{minipage}
\\
\begin{minipage}{0.48\textwidth}
\centering
\includegraphics[width=1.0\textwidth]{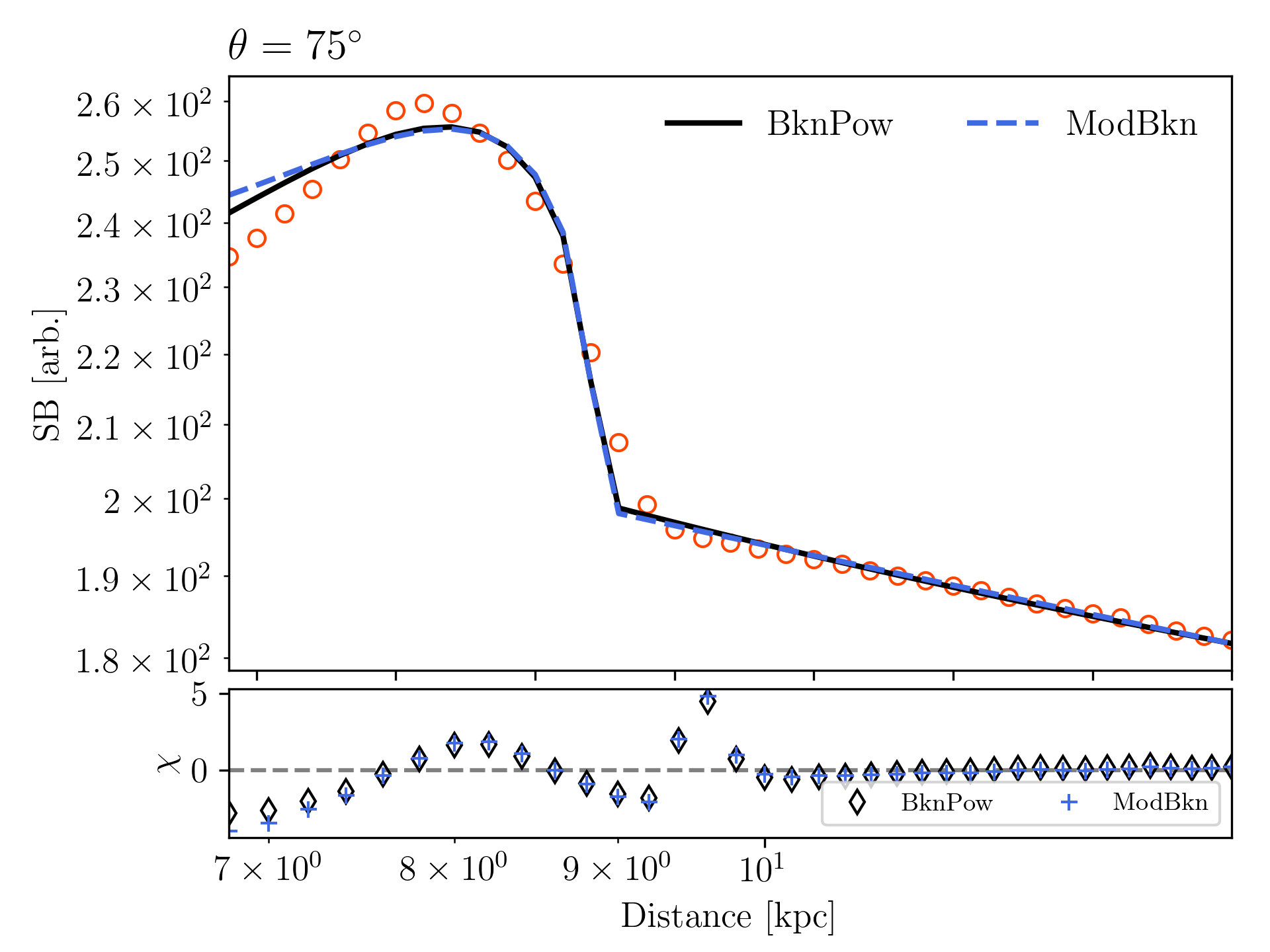}
\end{minipage} &

\begin{minipage}{0.48\textwidth}
\centering
\includegraphics[width=1.0\textwidth]{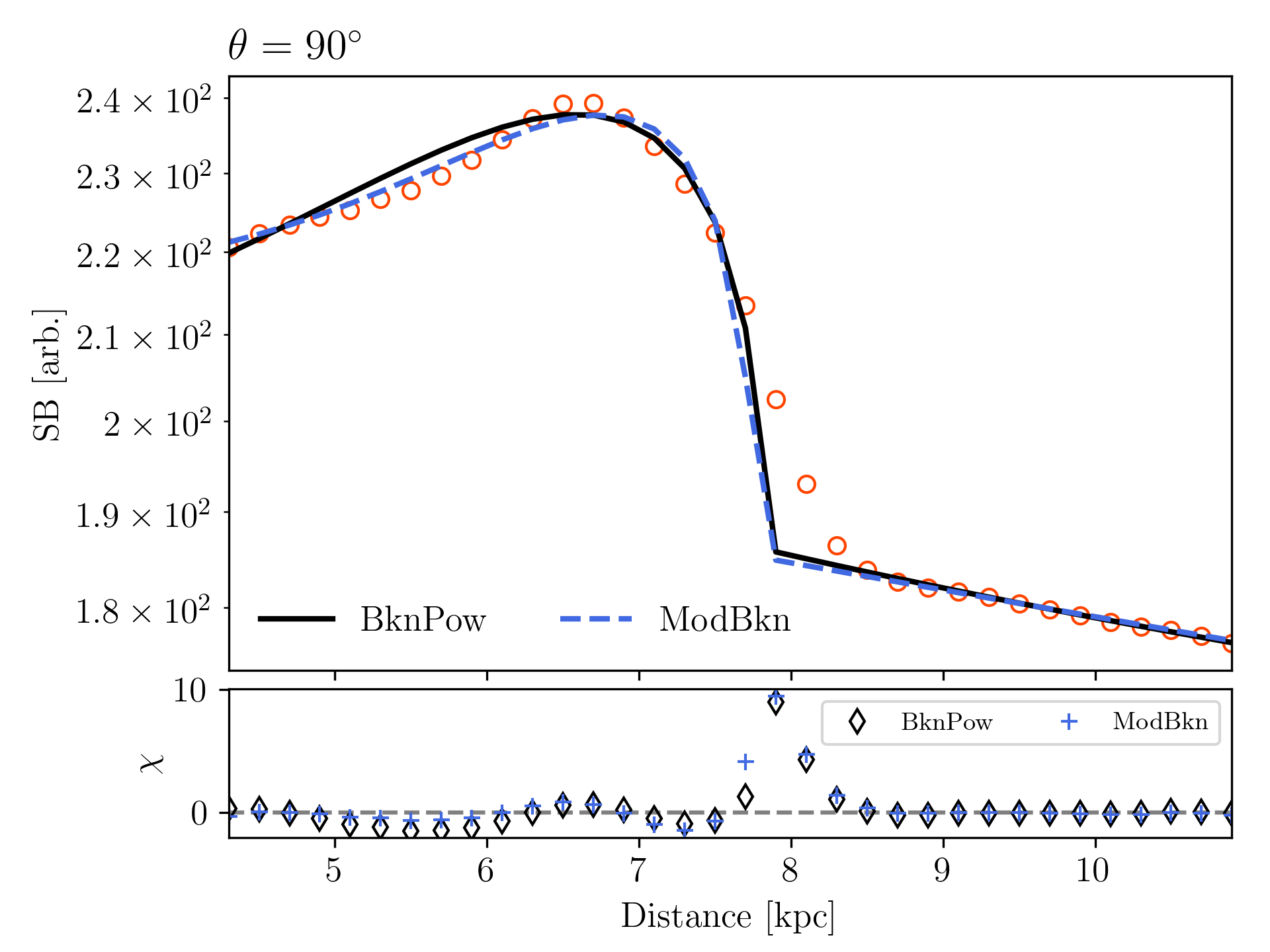}
\end{minipage}
\end{tabular}

\caption{Surface brightness profiles (red circle) across the forward shock and best fit results using {\tt BknPow} (black solid line) and {\tt ModBkn} models (Blue dotted line) at a 30 ({\bf top left}), 45 ({\bf top right}), 52 ({\bf middle left}), 60 ({\bf middle right}), 75 ({\bf bottom left}), and 90 ({\bf bottom right}) degree viewing angels.}
\label{fig_a3}
\end{figure*}

Figure \ref{fig_a1} and \ref{fig_a2} present simulated X-ray surface brightness maps and spectroscopic-like temperature maps for different viewing angles.
The surface brightness profile and the best fit of both models are shown in Figure \ref{fig_a3}.

%%------------------------------------------------------------%%
%%------------------------------------------------------------%%
%%------------------------------------------------------------%%
%%------------------------------------------------------------%%

%%-----------------------------------------------------------%%

\section{Best fitting of Cygnus A}
\label{sec:appendC}
\begin{figure*}
\begin{tabular}{cc}
\begin{minipage}{0.48\textwidth}
\centering
\includegraphics[width=1.0\textwidth]{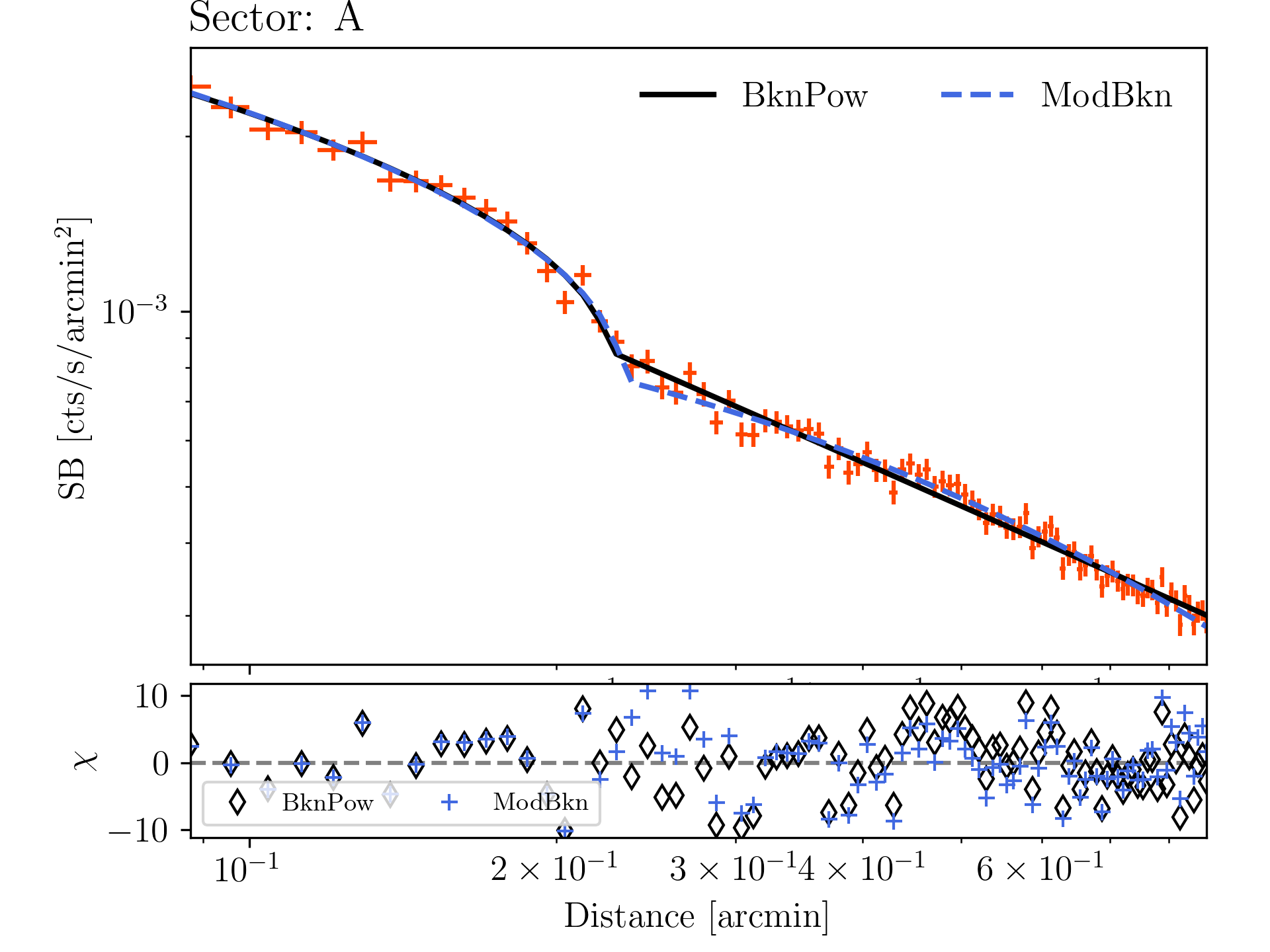}
\end{minipage} &

\begin{minipage}{0.48\textwidth}
\centering
\includegraphics[width=1.0\textwidth]{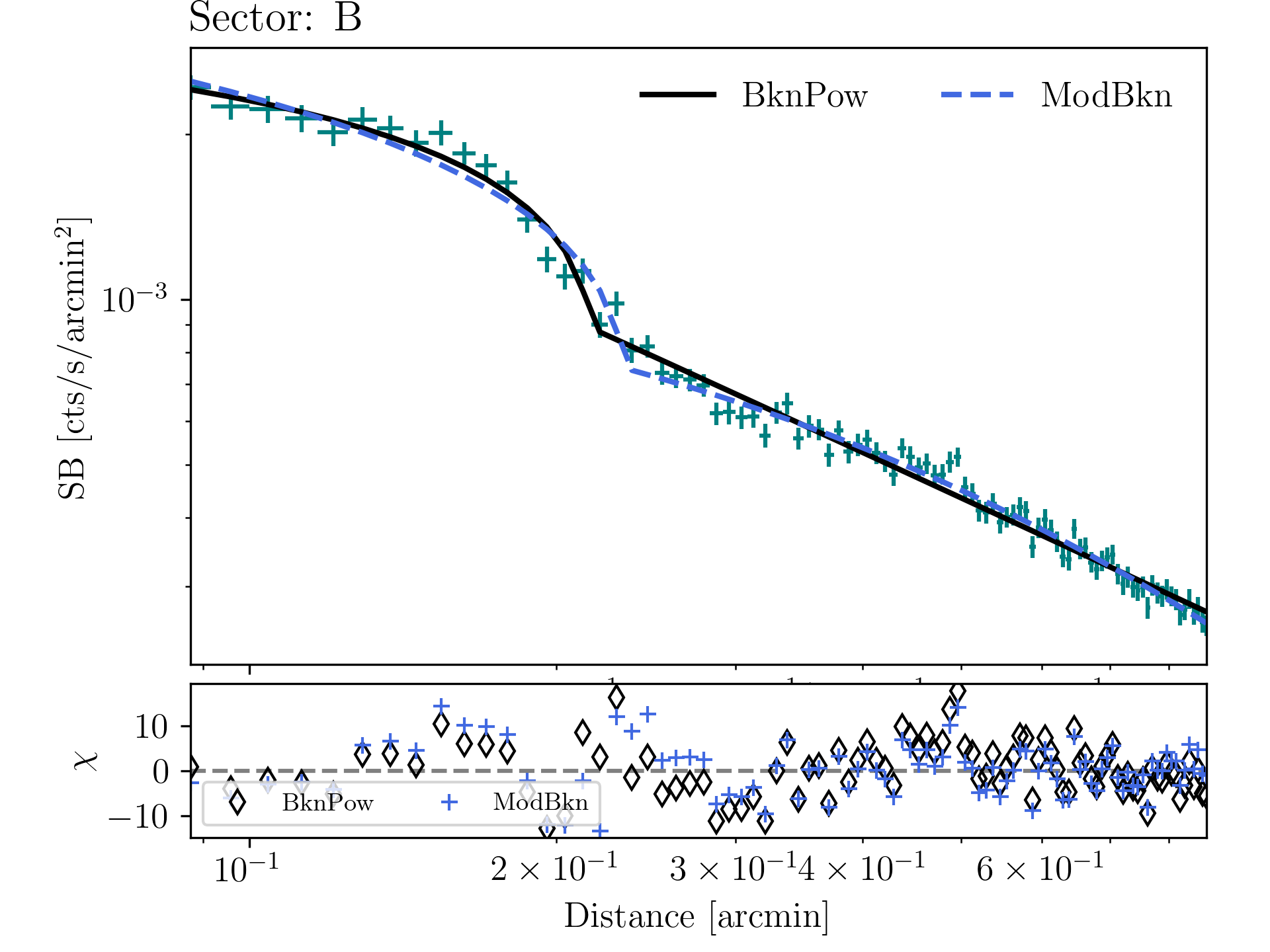}
\end{minipage}
\\
\begin{minipage}{0.48\textwidth}
\centering
\includegraphics[width=1.0\textwidth]{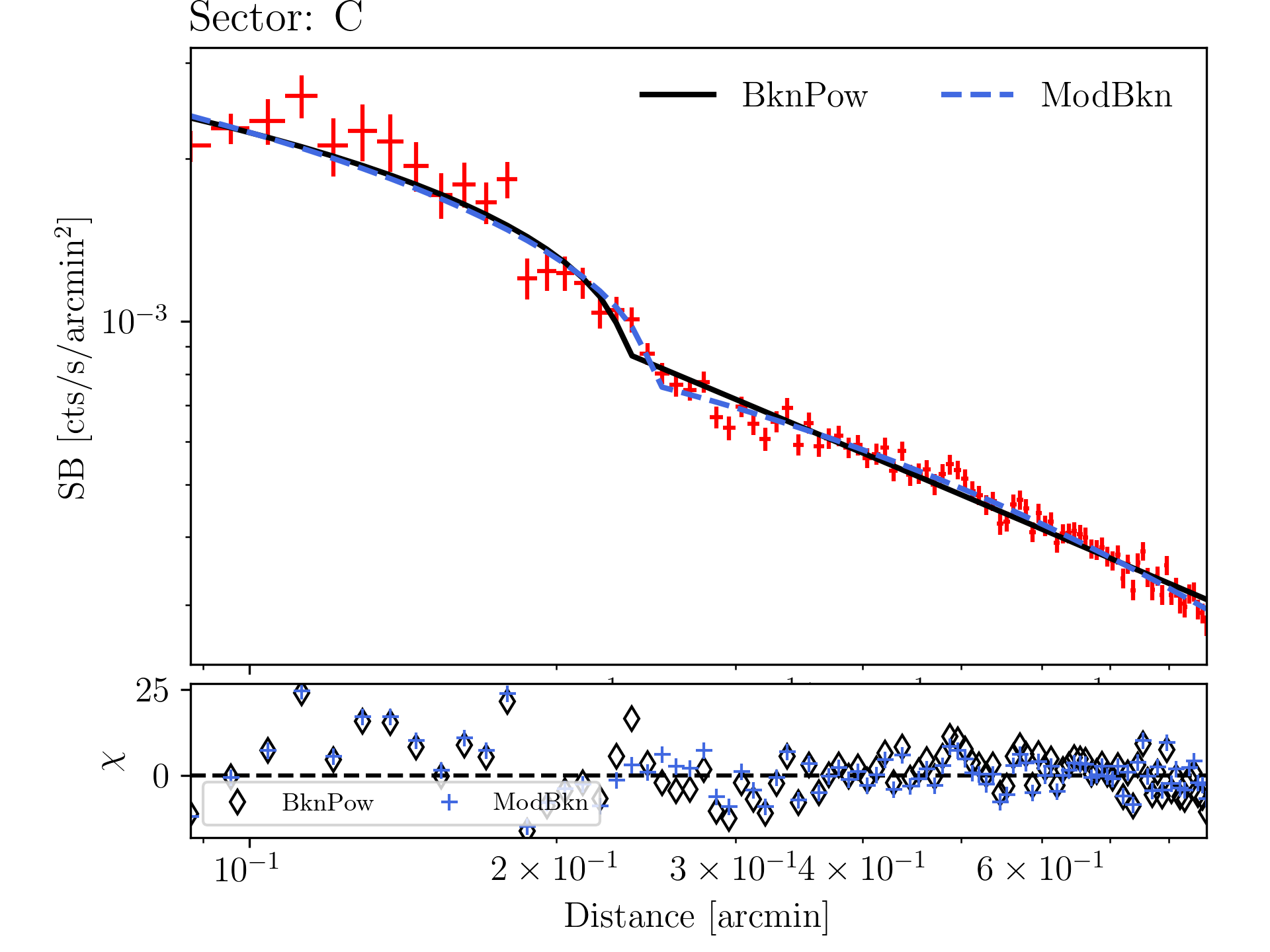}
\end{minipage} &

\begin{minipage}{0.48\textwidth}
\centering
\includegraphics[width=1.0\textwidth]{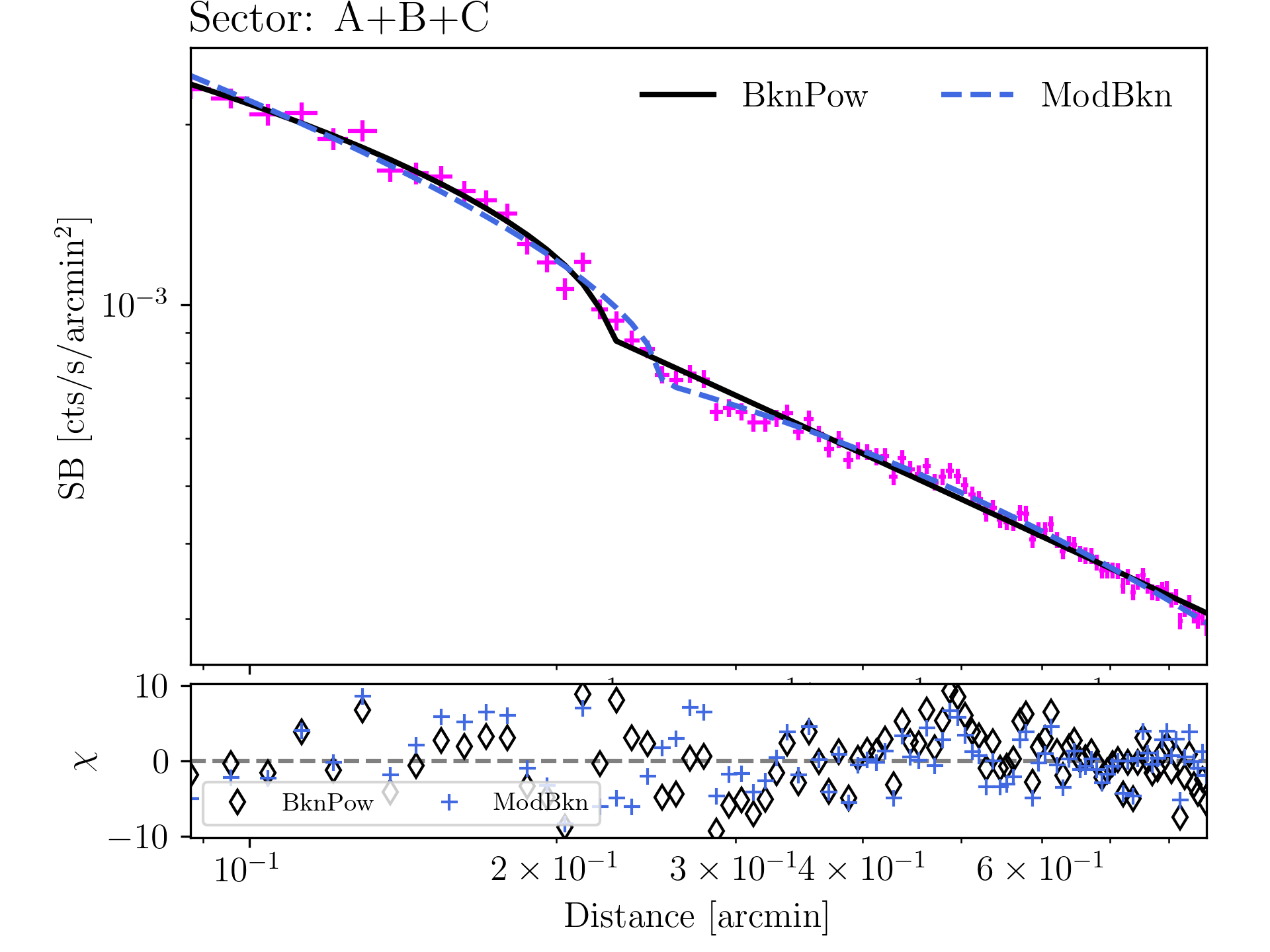}
\end{minipage}
\end{tabular}
\caption{Surface brightness profile (cross) across the forward shock of Cygnus A for Sector A ({\bf top left}), B ({\bf top right}), C ({\bf bottom left}) and A+B+C ({\bf bottom right}). 
Best fits results by using {\tt BknPow} (black solid line) and {\tt ModBkn} (Blue dotted line) models are over-plotted. }
\label{fig_b1}
\end{figure*}

Figure \ref{fig_b1} presents the radial profile and best fits of each sector, which are shown in Figure \ref{fig12}.

\end{appendix}


\end{document}